
\input epsf
%
\input harvmac
%
\font\twelverm=cmr12  \font\twelveit=cmti12
\font\twelvesl=cmsl12 \font\twelvebf=cmbx12
\font\twelvei=cmmi12  \skewchar\twelvei='177
\font\twelvesy=cmsy10 scaled\magstep1 \skewchar\twelvesy='60
\def\twelvepoint{\def\rm{\fam0\twelverm}
\textfont0=\twelverm \scriptfont0=\ninerm \scriptscriptfont0=\sixrm
\textfont1=\twelvei  \scriptfont1=\ninei  \scriptscriptfont1=\sixi
\textfont2=\twelvesy \scriptfont2=\ninesy \scriptscriptfont2=\sixsy
\textfont\itfam=\twelveit \def\it{\fam\itfam\twelveit}
\textfont\bffam=\twelvebf \def\bf{\fam\bffam\twelvebf}
\def\sl{\fam\slfam\twelvesl}\rm}
\def\secfont{\twelvepoint}
\def\newsec#1{\global\advance\secno by1\message{(\the\secno. #1)}
\global\subsecno=0\eqnres@t\centerline{\secfont \the\secno. #1}
\writetoca{{\secsym} {#1}}\par\nobreak\medskip\nobreak}
\def\eqnres@t{\xdef\secsym{\the\secno.}\global\meqno=1\bigbreak\bigskip}
\def\sequentialequations{\def\eqnres@t{\bigbreak}}\xdef\secsym{}
\def\appendix#1#2{\global\meqno=1\global\subsecno=0\xdef\secsym{\hbox{#1.}}
\bigbreak\bigskip\centerline{\secfont Appendix #1. #2}\message{(#1. #2)}
\writetoca{Appendix {#1.} {#2}}\par\nobreak\medskip\nobreak}
\def\listrefs{\footatend\bigskip\bigskip\immediate\closeout\rfile\writestoppt
\baselineskip=13pt\centerline{{\secfont References}}
\bigskip{\frenchspacing \parindent=20pt\escapechar=`
\input\jobname.refs\vfill\eject}\nonfrenchspacing}%
%
\def\figin{\epsfcheck\figin}\def\figins{\epsfcheck\figins}
\def\epsfcheck{\ifx\epsfbox\UnDeFiNeD
\message{(NO epsf.tex, FIGURES WILL BE IGNORED)}
\gdef\figin##1{\vskip2in}\gdef\figins##1{\hskip.5in}
\else\message{(FIGURES WILL BE INCLUDED)}%
\gdef\figin##1{##1}\gdef\figins##1{##1}\fi}
\def\DefWarn#1{}
\def\figinsert{\goodbreak\midinsert}
\def\ifigc#1#2#3{\DefWarn#1\xdef#1{\the\figno}
\writedef{#1\leftbracket \the\figno}
\figinsert\figin{\centerline{#3}}\medskip
\centerline{\vbox{\baselineskip\footskip
\centerline{\footnotefont{\bf Fig.~\the\figno:} #2}}}
\bigskip\endinsert\global\advance\figno by1}
\def\ifig#1#2#3{\DefWarn#1\xdef#1{\the\figno}
\writedef{#1\leftbracket \the\figno}
\figinsert\figin{\centerline{#3}}\medskip
\centerline{\vbox{\baselineskip\footskip
\advance\hsize by -1truein\noindent
\footnotefont{\bf Fig.~\the\figno:} #2}}
\bigskip\endinsert\global\advance\figno by1}
%
\def\Title#1#2{\rightline{#1}\ifx\answ\bigans\nopagenumbers\pageno0
\vskip0.5in
\else\pageno1\vskip.5in\fi \centerline{\titlefont #2}\vskip .3in}

\font\caps=cmcsc10
\noblackbox
\parskip=1.5mm
%
\def\npb#1#2#3{{\it Nucl. Phys.} {\bf B#1} (#2) #3 }
\def\anpb#1#2#3{{\bf B#1} (#2) #3 }

\def\prd#1#2#3{{\it Phys. Rev. } {\bf D#1} (#2) #3 }
\def\prl#1#2#3{{\it Phys. Rev. Lett.} {\bf #1} (#2) #3 }
\def\rmp#1#2#3{{\it Rev. Mod. Phys.} {\bf #1} (#2) #3 }

\def\pr#1#2#3{{\it Phys. Rev. } {\bf #1} (#2) #3 }

\def\jpm#1#2#3{{\it J. Phys. (Moscow)} {\bf #1} (#2) #3 }
\def\th#1{{\tt hep-th/#1}}
\def\ph#1{{\tt hep-ph/#1}}

\def\dj{\hbox{d\kern-0.347em \vrule width 0.3em height 1.252ex depth
-1.21ex \kern 0.051em}}

\def\d{{\rm d}}
\def\rO{{\rm O}\,}

\def\Tr{{\rm Tr\,}}

\def\ket{\rangle}
\def\bra{\langle}
\def\bket{\big\rangle}
\def\bbra{\big\langle}

\def\pvi{-\hskip -11pt \int}
%
            
 \def\CC{{\cal C}} \def\CF{{\cal F}} 
\def\CL{{\cal L}}   
\def\CB{{\cal B}}   \def\CT{{\cal T}}
\def\CM{{\cal M}} \def\CP{{\cal P}}
%
\lref\pert{C.~B. Thorn, \prd {20}{1979}{1934.}}
\lref\thorn{C.~B. Thorn, \prd {20}{1979}{1435.}}
\lref\dirac{P.~A.~M. Dirac, \rmp {21}{1949}{392.}}
\lref\wein{S. Weinberg, \pr {150}{1966}{1313.}}
\lref\tammd{I. Tamm, \jpm {9}{1945}{449\semi}
S.~M. Dancoff, \pr {78}{1950}{382\semi}
R.~J. Perry, A. Harindranath and K. Wilson, \prl {65}{1990}{2959.}}
\lref\rWilson{K. G. Wilson, \prd {10} {1974} {2445.}}
\lref\paulib{S. J. Brodsky and H.-C. Pauli,
{\it ``Light-Cone Quantization of Quantum Chromodynamics,''}
Lectures at the 30th Schladming Winter School in
Particle Physics, SLAC-PUB-5558 (1991).}
\lref\adjoint{S. Dalley and I.~R. Klebanov, \prd {47}{1993}{2517\semi}
G. Bhanot, K. Demeterfi and I.~R. Klebanov, \prd {48}{1993}{4980\semi}
K. Demeterfi, I.~R. Klebanov and G. Bhanot, \npb {418}{1994}{15.}}
\lref\largeN{G. 't Hooft, \npb {72}{1974}{461.}}
\lref\thooft{G. 't Hooft, \npb {75}{1974}{461.}}
\lref\hornbostel{ K. Hornbostel, {\it The Application of Light-Cone
Quantization to Quantum Chromodynamics in (1+1) Dimensions,}
Ph.D. thesis, SLAC report No. 333 (1988).}
\lref\durgut{M. Durgut, \npb {116}{1976}{233.}}
\lref\witten{E. Witten, \npb {160}{1979}{57.}}
\lref\einh{M.~B. Einhorn, \prd {14}{1976}{3451.}}
\lref\gross{C.~G. Callan, N. Coote and D.~J. Gross,
\prd {13}{1976}{1649.}}
\lref\jaffe{R.~L. Jaffe and P.~F. Mende, \npb {369}{1992}{189.}}
\lref\gm{B. Grinstein and P.~F. Mende, {\it ``Exact Heavy to Light
Meson Form Factors in the Combined Heavy Quark, Large $N_c$ and
Chiral Limits,''} Preprint Brown HET-928, \ph{9312353.}}
\lref\gmm{B. Grinstein and P.~F. Mende, {\it ``Form Factors in the
Heavy Quark and Chiral Limit: Pole Dominance in
$\bar{B}\to\pi e\bar{\nu}_e$,''}
Preprint Brown HET-930, SSCL-549, \ph{9401303.}}
\lref\collect{E. Br\'ezin and S.~R. Wadia, {\it The Large N Expansion
in Quantum Field Theory and Statistical Physics -- From Spin Systems
to 2--dimensional Gravity,} (World Scientific, Singapore, 1993).}
\lref\bglcg{I. Bars and M.~B. Green, \prd {17}{1978}{537.}}
\lref\bars{I. Bars and A.~J. Hanson, \prd {13}{1976}{1744.}}
\lref\numer{A.~J. Hanson, R.~D. Peccei and M.~K. Prasad,
\npb {121}{1977}{477.}}
\lref\grosstay{D.~J. Gross, \npb {400}{1993}{161\semi}
D.~J. Gross and W. Taylor, \npb {400}{1993}{181;}
\anpb {403}{1993}{395.}}
%

\line{\hfill PUPT-1461}
\vskip 0.5cm

\Title{\vbox{\baselineskip 12pt\hbox{}
 }}
{\vbox{
\centerline{Effective hamiltonians for $1/N$ expansion}
\vskip0.2in
\centerline{in two-dimensional QCD}
 }}

\vskip 0.5cm

\centerline{{\caps J. L. F. Barb\'on}\footnote{$^{1}$}{E-mail:
barbon@puhep1.princeton.edu}
and
{\caps Kre\v simir Demeterfi}\footnote{$^{2}$}
{On leave of absence from the Ru\dj er Bo\v skovi\'c Institute,
Zagreb, Croatia}{$^{,}$}\footnote{$^{3}$}{E-mail:
kresimir@puhep1.princeton.edu} }
\smallskip
\centerline{{\sl Joseph Henry Laboratories}}
\centerline{{\sl Princeton University}}
\centerline{{\sl Princeton, NJ 08544, U.S.A.}}
\vskip 0.8in
%
%
We discuss the general structure of effective hamiltonians for
systematic $1/N$ expansion in QCD using the light-cone quantization.
These are second-quantized hamiltonians acting on the Fock space of
mesons and glueballs defined by the solution of the $N=\infty$ problem.
In the two-dimensional case we find only cubic and quartic
interaction terms, and give explicit expressions for the vertex
functions as integrals of solutions of 't Hooft equation.
As examples of possible applications of our formalism,
we study $1/N$ corrections to meson mass and form factors
for decays  of
$Q\bar{q}$ states, recently discussed by Grinstein and Mende in the
large-$N$ limit. We find that $1/N$ is a good small expansion
parameter.

\Date{6/94}

\newsec{Introduction}

It is generally believed that quantum chromodynamics (QCD) is the
correct theory of strong interactions. Because of the asymptotic
freedom one can use perturbation theory to explain
short-distance properties of hadrons.
At long distances, however,
the theory becomes strongly coupled exhibiting
color confinement, and nonperturbative treatment is clearly required.

As originally proposed by 't Hooft~\refs\largeN, generalizing the
gauge group of QCD to SU$(N)$ provides $1/N$ as the unique
small expansion parameter if $N$ is large (in reality, $N=3$).
It is hoped that in the limit $N\to\infty$
 the problem simplifies enough to become tractable (while still
capturing the essential dynamics), and that a systematic
$1/N$ expansion can be developed around such a solution.
Large-$N$ expansion has indeed been a very fruitful idea
in various branches of physics (see, c.f., ref.~\refs\collect).
In the physically most interesting case of QCD, however, this
approach has been successfully applied only in two space-time
dimensions, where 't Hooft solved the $N=\infty$ problem
for mesons by summing an infinite set of planar diagrams~\refs\thooft.
The solution
is given in the form of an integral equation for meson spectrum and
wave functions. In this approximation one finds
an infinite tower of stable mesons with
asymptotically linearly growing mass.
This model was further studied by Callan, Coote and Gross~\refs\gross,
and by Einhorn~\refs\einh\ who showed that the scattering amplitudes
and the electromagnetic form factors are given completely in terms of
asymptotic color-singlet bound states, and that no free quarks
appear in the physical spectrum.
More recently, Grinstein and Mende~\refs\gmm\ extended the analysis of
form factors to the case of flavor changing currents
and used the 't Hooft model to explicitly illustrate their earlier
argument that semileptonic $B$ decays in the four-dimensional
theory are dominated by a single pole
in the combined limits $N\to\infty, M_b\to\infty$ and
$m_\pi\to 0$ \refs\gm.

In this paper we discuss effective hamiltonians for systematic
$1/N$ expansion in QCD using the light-cone quantization.
The advantage of using light-cone quantization comes from the fact that
the  light-cone vacuum is an exact eigenstate of the full hamiltonian.
As a result, light-cone Fock space methods become powerful tools in
the study of relativistic bound-state problems~\refs\paulib.
A nice feature of the large-$N$ limit in this formulation is that the
corresponding bound-state equations become linear equations for
the so-called light-cone wave functions\footnote{$^*$}{The Fock space
formulation of the $1/N$ expansion
of QCD has been advocated some time ago by C.~Thorn~\refs\thorn.}.
The main idea of our approach is to construct
a second-quantized hamiltonian acting on the Fock space of
mesons and glueballs defined by the solution of the $N=\infty$
problem. As usual,
this hamiltonian is defined by requiring the equivalence
of its matrix elements between the Fock space states to the matrix
elements of the fundamental QCD hamiltonian between asymptotic states.
This effectively defines the vertex functions
as nested commutators of the light-cone hamiltonian with the operators
creating asymptotic states. These commutators can be expressed
solely in terms of the solutions of the large-$N$ eigenvalue problem.
Because of the structure of the light-cone hamiltonian and the
light-cone Fock space, such effective hamiltonian will have only a
{\it finite} number of interaction terms, but with non-local vertices.
Once the vertex functions are known one can
develop a systematic $1/N$ expansion
by using the standard light-cone perturbation theory.
Our method is quite general
and can be easily extended to the baryon sector, as
well as to higher dimensions, provided the planar solution
is known.

We carry out this program explicitly for the meson sector
of two-dimensional QCD. Using
the hamiltonian light-cone formalism is particularly useful
in this case because it eliminates the need to  sum gluon diagrams
which are summarized in the effective Coulomb interaction
by solving the gauge field constraint.
We find only cubic and quartic interactions and derive
expressions for the vertex functions in terms of
't Hooft functions.
Since solutions of 't Hooft equation are not known in terms of
standard functions, we can only give numerical estimates of the
vertex functions. We present several examples of these
vertex functions obtained numerically and discuss
their properties relevant for the $1/N$ expansion.
Using the standard light-cone perturbation theory
we study the $1/N$ corrections to
the meson spectrum and to the $B$-meson decay form factors,
and find that $1/N$ for $N=3$ is indeed effectively a
small parameter.

The outline of our presentation goes as follows. In sect.~2 we
discuss the general structure of effective hamiltonians for the
$1/N$ expansion in QCD, independent of the dimension of space-time.
In sect.~3 we review the light-cone quantization of
two-dimensional QCD, and write the full light-cone hamiltonian
in terms of color-singlet operators. In sect.~4 we rederive the
bound-state equations for mesons and baryons in the large-$N$
limit in order to demonstrate power of the hamiltonian light-cone
formalism which does not require summing diagrams.
In sect.~5 we present the main result of our paper -- construction
of the effective hamiltonian for mesons in two-dimensional QCD.
As examples of possible applications of our formalism
we study $1/N$ corrections to the meson spectrum in sect.~6
and to the form factors in sect.~7.
Finally, in sect.~8 we make some comments on the analytic
properties, and end with conclusions and a few general
remarks in sect.~9.
In the appendix we collect explicit expressions for vertex functions.

\newsec{Effective hamiltonians}

Given the $N=\infty$ solution of QCD, it  would be easy
to rewrite the $1/N$ expansion in a formal hamiltonian framework.
The planar solution is known to be a free bosonic Fock space of mesons
and glueballs. Let us denote their creation operator generically by
$A^{\dagger}$. The asymptotic Hilbert space at $N=\infty$ is then
spanned by states of the form
\eqn\focks
{A_{i_1}^{\dagger}\dots A_{i_n}^{\dagger}|0\ket\,\,,\qquad
[A_i , A_{j}^{\dagger} ] = \delta_{ij}\,\,,}
where the indices stand for the whole set of meson/glueball quantum
numbers.

The perturbative physics in powers of $1/N$ depends on the matrix elements,
\eqn\matel
{\bra  0 | A_{j_1}\dots A_{j_q} H A_{i_1}^{\dagger}\dots
A_{i_p}^{\dagger} |0 \ket\,\,, }
where $H$ is the full QCD hamiltonian. An equivalent
second quantized hamiltonian
acting on the asymptotic Hilbert space \focks\ is easily constructed
by matching the matrix elements \matel. The solution is,
\eqn\effham
{H_{\rm eff} = \sum_{p,q} {1\over p! q!}\, H_{(p,q)}
(j_1,\dots,j_q | i_1,..., i_p) A_{j_1}^{\dagger}\dots A_{j_q}^{\dagger}
A_{i_1}\dots A_{i_p}\,\,, }
with irreducible many-body interactions given by the matrix elements,
\eqn\nested
{H_{(p,q)} (j_1,\dots, j_q | i_1,\dots i_p) =\bbra 0 \bigl|
A_{j_1}\dots A_{j_q} \bigl[\dots \bigl[ H, A_{i_1}^{\dagger}\bigr]\dots
A_{i_p}^{\dagger} \bigr] \bigr|0\bket\,\,. }
These quantities should be calculated from the fundamental quark-gluon
theory. In particular, we must know the expressions for
$A_{i}^{\dagger}$ in terms of the fundamental variables (wave
functions), which in turn is equivalent to diagonalization of
the $H_{(1,1)}$ part of the effective hamiltonian (the
planar, $N=\infty$, problem). Of course, this is a formidable
task in practice. As pointed out by Thorn~\refs\thorn, such  Fock
space representation of the $1/N$ expansion is very involved in the
standard equal-time quantization, although it
simplifies considerably if we adopt the light-cone quantization
scheme~\refs\dirac\ or, in more physical terms, if we look at the
system from the infinite momentum frame~\refs\wein.

Let us briefly review Thorn's argument. The main simplifying
feature of the light-cone hamiltonians is the absence of
pure positive/negative frequency terms, with the result that
the Fock vacuum becomes an exact eigenstate
of the full (interacting) hamiltonian. As far as $1/N$
power-counting is concerned we can forget about all quantum numbers
except color. The  SU$(N)$ structure for the glue sector of QCD is
\eqn\tham
{H_{LC} \sim {\rm Tr}( a^{\dagger} a) + {1\over \sqrt{N}}{\rm Tr}
(a^{\dagger} a^{\dagger} a) +  {1\over N}{\rm Tr}( a^{\dagger}
a^{\dagger} a a )+ {1\over N}\Tr (a^{\dagger} a a^{\dagger} a )
+ {\rm h.c.}     }
where $a^{\dagger}$ is an $N\times N$ operator-valued hermitian
matrix of transverse gluon creators. It is then easy to see
that the $N=\infty$ eigenvalue equation closes on the subspace of
invariant states of the form
\eqn\strings
{G^{\dagger} |0\ket =\sum_l f(1,\dots,l) {1\over N^{l/2}} {\rm Tr}
(a_1^{\dagger}\dots a_l^{\dagger} ) |0\ket\,\,,  }
which can be represented graphically as a superposition of closed
strings each containing $l$ ``partons". The coefficients $f(1,\dots,l)$
are usually referred to as the light-cone wave functions.
In fact, the complete set of ``string"
states \strings\ provides a plane-wave basis for Wilson loop
operators, and the glueball states are clearly built directly
in loop space.

In the equal-time quantization, in addition to terms in eq.~\tham,
the hamiltonian contains pure positive frequency pieces, like
${1\over \sqrt{N}}{\rm Tr}( a^{\dagger} a^{\dagger} a^{\dagger })$.
When these terms are included, the $N=\infty$ problem
closes on a much more complicated space of coherent states
of invariant traces (previous string-like states).
A simple 0-dimensional toy model from ref.~\thorn\ illustrates this
fact. The hamiltonian
\eqn\toy
{H = \epsilon_0 \Tr (a^{\dagger} a)+{1\over \sqrt{N}}\,
\Bigl(\Tr(a^{\dagger} a^{\dagger} a^{\dagger} ) + {\rm h.c.}\Bigr)}
has eigenvalue $E= -N^2 \epsilon_0 $ for the eigenstate
$$
|\psi\ket = {\rm exp}\left\{ -{N\over 3\epsilon_0} \Tr\left(
{a^{\dagger} \over \sqrt{N}}\right)^3 \right\} |0\ket \,\,.  $$

The form of the coherent states of loops for the real problem \tham\
is not known, and the large-$N$ Fock space is  too big
even for numerical approach.  As a result, the
only hamiltonian Fock space method that could work in practice
is the light-cone one.

Using eqs.~\nested, \tham\ and \strings\ one easily finds the following
structure of the effective light-cone hamiltonian in the glueball
sector:
\eqn\glueham
{H_{\rm eff} \sim H_{(1,1)} G^{\dagger} G + {1
\over 2 N} H_{(1,2)} G^{\dagger} G^{\dagger} G  + {1\over 4 N^2}
H_{(2,2)} G^{\dagger} G^{
\dagger} G G  + {1\over 6 N^2} H_{(1,3)} G^{\dagger} G^{\dagger}
G^{\dagger} G   + {\rm h.c.}}
The planar hamiltonian is defined by the one-particle sector in
this expansion, $H_0 = H_{(1,1)} G^{\dagger} G$, and its
diagonalization constitutes the large-$N$ solution of the
model. In general, $H_{(1,1)}$ is a complicated integral kernel.
For example, for fields in the adjoint representation it is given
by an infinite hierarchy of integral operators, accounting
for the interaction of partons inside strings \strings\
of arbitrary length~\refs\adjoint.
The working hypothesis in this whole approach is that only a small
number of partons contributes significantly to the low-lying states.
This goes under the name of light-cone Tamm-Dancoff
approximation~\refs\tammd\ and numerical approaches, like the
discretized light-cone method~\refs\paulib, automatically incorporate
such Fock space truncations. Encouraging evidence for the validity of
this procedure is found in the analysis of simple two-dimensional
models with complicated loop space, like two-dimensional QCD with
adjoint matter~\refs\adjoint.

The actual computation of the kernels \nested\ involves numerical
analysis even in the simplest possible case, the 't Hooft model.
However, an interesting general feature of the light-cone effective
hamiltonians \effham\ is that they contain only a finite number of
vertices, albeit non-local ones. These non-local vertices replace the
infinite tower of higher-derivative operators in usual local
expansions of effective lagrangians. The relation between our
light-cone hamiltonians and the standard covariant effective
lagrangians is very much the same as the relation between light-cone
string field theory and covariant string field theory: the former has
only cubic interactions while the latter has a non-polynomial action.

As a final remark, we point out that  by construction \nested\ only
depends on the meson/glueball quantum numbers, while the fundamental
hamiltonian $H$ depends on all quark/gluon quantum numbers. Thus,
deriving \nested\ involves ``integrating out" part of the
microscopic degrees
of freedom, justifying the use of the word ``effective". However, it
is important to note that no low-energy expansion is assumed, at least
in principle, as long as the large-$N$ problem is solved exactly. The
nature of the approximation is more similar to the Born-Oppenheimer
method, where fast and slow variables are decoupled at the level of
wave functions. The effective hamiltonians considered here give the
exact answer for perturbative questions about the free asymptotic Fock
space, like $1/N$ corrections to scattering amplitudes and
self-energies, but their description of ``nuclear physics" is only
approximate. For example, exotics would be treated as meson bound
states, rather than four-quark bound states, in exact analogy with the
Born-Oppenheimer decoupling in molecular physics.

In the remainder of this paper we study in detail the effective
light-cone hamiltonian for two-dimensional QCD. In this case one can
integrate out the gluons by solving the light-cone gauge constraint,
and the large-$N$ Fock space has a simple structure encoded in the solutions
of 't Hooft's bound-state equation. This model illustrates many of the
issues involved in the large-$N$ approach
and the structure we find is very similar to the one
of four-dimensional QCD in the valence approximation (constituent
quark model).

\newsec{The two-dimensional model}

In this section we review the light-cone quantization of
two-dimensional QCD with $N_f$ flavors of quarks in the fundamental
representation of SU$(N)$, and fix the notation to be used
throughout the paper. The model is given by the lagrangian,
\eqn\lagr{
\CL=-{1\over 4}\,\Tr F^2 +\sum_{j=1}^{N_f}\,
\bar{\psi}_j(i\Dsl -m_j)\psi_j\,\,,}
where
$iD_\mu=i\partial_\mu-gA_\mu^a T^a$,
and the SU($N$) generators, $T^a$, are normalized as
\eqna\normT
$$\eqalignno{
&\Tr(T^aT^b)=\delta^{ab}\,\,,&\normT {\rm a}\cr
\noalign{\vskip 0.2cm}
&\sum_a\,T^a_{\alpha\beta}T^a_{\gamma\delta}=
\delta_{\alpha\delta}\delta_{\beta\gamma}-{1\over N}\,
\delta_{\alpha\beta}\delta_{\gamma\delta}\,\,.&\normT {\rm b}\cr}$$
We shall work in the light-cone
coordinates $x^{\pm}=(x^0\pm x^1)/\sqrt{2}$, and
use the following representation for $\gamma$-matrices:
$\gamma^0=\sigma_2, \gamma^1=i\sigma_1$.
The light-cone hamiltonian, $P^{-}$, and
momentum, $P^{+}$, are given by
\eqn\enmom{
P^{\mp}=\int\limits_{x^+={\rm const.}}T^{+\mp}\,\,,}
where $T^{\mu\nu}$ is the energy-momentum tensor.
In the light-cone quantization one takes
$x^+={\rm const.}$ as initial value surfaces for
the light-cone hamiltonian.

In the studies of gauge theories on the light cone,
it is most convenient to work in the
light-cone gauge, $A_{-}=0$, which introduces the following
constraints:
\eqn\constraints{ \eqalign{
&i\gamma^{-}\partial_{-}\psi_j={1\over 2}\,
m_j\gamma^{-}\gamma^{+}\psi_j\,,\cr
\noalign{\vskip 0.2cm}
&(i\partial_{-})^2 A_{+}^{a}=g\sum_{j=1}^{N_f}
\bar{\psi}_{j}\gamma^{+}T^{a}\psi_j\,\,,\cr
\noalign{\vskip 0.2cm}
&\sum_{j=1}^{N_f}\,\,\, \int\limits_{x^+={\rm const.}}
\bar{\psi}_{j}\gamma^{+}T^{a}\psi_j\,|\,{\rm  phys}\,\ket =0\,\, .}}
The first constraint projects out a component of the quark field and
it is purely kinematical.
The solution of the second constraint,
\eqn\Aplus{
A_{+}^{a}=g\sum_{j=1}^{N_f}\,{1\over (i\partial_{-})^2}\,
\bar{\psi}_{j}\gamma^{+}T^{a}\psi_j\,\,,}
induces the well-known Coulomb potential between the fermion currents,
which is responsible for all the static properties of the model.
The zero-mode of the gauge field $A_+^a$ does not enter the second
constraint. Direct integration over this zero-mode leads to the third
constraint which restricts the Fock space states to color singlets.
The last two constraints therefore entail the confinement property of
the model.
After solving constraints \constraints\ the light-cone hamiltonian reads,
\eqn\lcham{
P^{-}=\int\limits_{x^+={\rm const.}}\,\,\biggl(
\sum_{j}\,{m_{j}^2\over\sqrt{2}}\, \psi_{j}^{\dagger}\,
{1\over i\partial_{-}}\,\psi_{j} +
g^2\,\sum_{j,k}\,(\psi_{j}^{\dagger}T^{a}\psi_{j})\,
{1\over (i\partial_{-})^2}\,(\psi_{k}^{\dagger}T^{a}\psi_{k})
\biggr)\,\,,}
and the light-cone momentum is given by,
\eqn\lcmom{
P^+= {i\over\sqrt{2}}\,\sum_{j}\int\limits_{x^+={\rm const.}}
\psi_{j}^{\dagger}{\buildrel\leftrightarrow\over\partial}_{-}
\,\psi_{j}\,\,.}
In eqs.~\lcham\ and \lcmom\
$\psi_{j}$ represents physical degree of freedom --
the projected out fermion field:
$\psi_{j}\to {1\over 2}\,\gamma^{-}\gamma^{+}\psi_{j}$.

In order to exhibit the $1/N$ power counting, it is convenient to
rewrite $P^{-}$ in a Fock space representation. This is
straightforward upon introducing the mode expansion:
\eqn\modeexp{
\psi_{j,\alpha}={1\over 2^{1/4}}\,\int_0^\infty
{\d k^{+}\over\sqrt{2\pi}}\,\Bigl(b_{j,\alpha}(k^+)\,\e{-ik^{+}x^{-}}
+d_{j,\alpha}^{\dagger}(k^+)\,\e{ik^{+}x^{-}}\Bigr)\,\,,}
where $\alpha$ is a color index, and the operators $b$ and $d$
satisfy canonical anticommutation relations:
\eqn\cac{
\{b_{i,\alpha}(k^+),b^{\dagger}_{j,\beta}(p^+)\}=
\{d_{i,\alpha}(k^+),d^{\dagger}_{j,\beta}(p^+)\}=
\delta_{\alpha\beta}\,\delta_{ij}\,\delta(k^{+}-p^{+})\,\,.}
(The normalization of $\psi$ is chosen to close the Poincar{\'e}
algebra.)
To simplify notation we introduce a collective index
$k\equiv(k^{+},i)$ to denote both momentum $k^{+}$ and the
flavor index $i$. In this notation
$$\int \d k \equiv \sum_{i=1}^{N_f}\,
\int\limits_0^\infty \d k^{+}\,\,.$$

After normal ordering the fermion bilinears,
the light-cone hamiltonian \lcham\ can be written
in the creation--annihilation basis in terms of
color-singlet operators:
\eqn\singlets{\eqalign{
&M^{\dagger}_{kk'}\equiv\sqrt{N}\,\sum_{\alpha}\,
b^{\dagger}_{\alpha}(k)\,d^{\dagger}_{\alpha}(k')\,\,,\cr
&M_{kk'}\equiv\bigl(M^{\dagger}_{kk'}\bigr)^{\dagger}\,\,,\cr
&B_{kk'}\equiv\sum_{\alpha}\,b^{\dagger}_{\alpha}(k)\,b_{\alpha}(k')\,\,,\cr
&D_{kk'}\equiv\sum_{\alpha}\,d^{\dagger}_{\alpha}(k)\,d_{\alpha}(k')\,\,.}}
These operators have obvious physical meaning. For example,
$M^{\dagger}_{kk'}$ ($M_{kk'}$) creates (annihilates) a meson in
which quark of flavor $i$ carries momentum $k^+$, while the antiquark
of flavor $i'$ carries momentum $k'^+$. Similarly,
$B_{kk'}$ ($D_{kk'}$) counts the number of quarks (antiquarks).
It is convenient to write $P^-$ as a sum of three terms,
\eqn\hamsplit{
P^-=T+V+V_{\rm sing}\,\,.}
The kinetic energy, $T$, is given by
\eqn\kin{
T=\sum_{j=1}^{N_f}\,{\bar{m}_j^2\over 2}\,\int {\d k^{+}\over k^{+}}\,
\bigl(B_{kk}+D_{kk}\bigr)\,\,,}
with the renormalized mass
\eqn\massr{
\bar{m}_j^2=m_j^2-{g^2 N\over\pi}\Bigl(1-{1\over 2N^2}\Bigr)\,\,.}
The potential energy, $V$, is given by
\eqn\pot{\eqalign{
V=&-{g^2N\over 4\pi}\,\int\d k\,\d k'\,\d p\,\d p' \,
\Bigl({\cal K}_{\rm MM}\,M^\dagger_{kk'}M_{pp'}
+{\cal K}_{\rm MB}\,M^\dagger_{kk'}\,B_{pp'} \cr
\noalign{\vskip 0.2cm}
&+{\cal K}_{\rm MD}\,M^\dagger_{kk'}\,D_{pp'}
+{\cal K}_{\rm BD}\,B_{kk'}\,D_{pp'}
+{\cal K}_{\rm BB}\,B_{kk'}\,B_{pp'}\cr
\noalign{\vskip 0.2cm}
&+{\cal K}_{\rm DD}\,D_{kk'}\,D_{pp'}+{\rm h.~c.} \Bigr)\,\,,}}
with the kernels given explicitly by:
\eqn\kernels{\eqalign{
&{\cal K}_{\rm MM} =\left[
{\delta_{ij}\delta_{i'j'}\over (k^{+}-p^{+})^2}
+{1\over N}\,{\delta_{ii'}\delta_{jj'}\over (k^{+}+k'^{+})^2}
\right]\,\delta(k^{+}+k'^{+}-p^{+}-p'^{+})\,\,,\cr
\noalign{\vskip 0.2cm}
&{\cal K}_{\rm BB} = {\cal K}_{\rm DD} = {1\over 2N}\,\left[
{\delta_{ij'}\delta_{ji'}\over (k^{+}-p'^{+})^2}
+{1\over N}\,{\delta_{ii'}\delta_{jj'}\over (k^{+}-k'^{+})^2}
\right]\,\delta(k^{+}-k'^{+}+p^{+}-p'^{+})\,\,,\cr
\noalign{\vskip 0.2cm}
&{\cal K}_{\rm MB} ={2\over \sqrt{N}}\,\left[
{\delta_{ij'}\delta_{ji'}\over (k^{+}-p'^{+})^2}
+{1\over N}\,{\delta_{ii'}\delta_{jj'}\over (k^{+}+k'^{+})^2}
\right]\,\delta(k^{+}+k'^{+}+p^{+}-p'^{+})\,\,,\cr
\noalign{\vskip 0.2cm}
&{\cal K}_{\rm MD} = -{2\over \sqrt{N}}\,\left[
{\delta_{ij}\delta_{i'j'}\over (k^{+}+p^{+})^2}
+{1\over N}\,{\delta_{ii'}\delta_{jj'}\over (k^{+}+k'^{+})^2}
\right]\,\delta(k^{+}+k'^{+}+p^{+}-p'^{+})\,\,,\cr
\noalign{\vskip 0.2cm}
&{\cal K}_{\rm BD} =-{1\over N}\,\left[
{\delta_{ij}\delta_{i'j'}\over (k^{+}+p^{+})^2}
+{1\over N}\,{\delta_{ii'}\delta_{jj'}\over (k^{+}-k'^{+})^2}
\right]\,\delta(k^{+}-k'^{+}+p^{+}-p'^{+})\,\,.\cr }}
The singular term, $V_{\rm sing}$, reads
\eqn\sing{
V_{\rm sing}={g^2 N\over 4\pi}\,{N_f\over N}\,
\left( \int\d k_1^+\,\d k_2^+\,
{\delta(k_1^+ - k_2^+)\over (k_1^+ - k_2^+)^2}\right)\,
\int \d k\, \bigl(B_{kk}+D_{kk}\bigr) + \rO(N^2)\,\,.}
The constant term $\rO(N^2)$, which we did not write down explicitly,
is an infinite vacuum energy and appears
as a result of our choice of normal ordering.
Had we started with the completely normal-ordered hamiltonian,
$:P^{-}:$, we would obtain a finite result with unshifted mass
$(\bar{m}_j=m_j)$. In fact, the precise definition of the quark mass
(being the mass of a colored object) is a matter of convenience.
In the next section we will comment more on this issue.
The singular term \sing\ arises because of our convention to write
$BB$ and $DD$ terms in \pot\ as color singlet operators,
and it cancels against similar terms in $V$ when we
compute matrix elements between meson states.
This completes the definition of the model which we study in the rest
of this paper.

\newsec{The 't Hooft equation}

In this section we summarize the large-$N$ solution of
two-dimensional QCD found by
't Hooft~\refs\thooft . We emphasize the power of the light-cone
method in deriving bound-state equations, since no diagrams need to
be considered. In fact, all gluon diagrams are summed over by means of
the effective Coulomb interaction \Aplus\ (see ref.~\refs\hornbostel ).

According to $1/N$ power counting in eq.~\kernels, the dominant
term in the
interaction hamiltonian corresponds to the $MM$ interaction,
\eqn\Vmm
{V_{\rm MM}=-{g^2 N\over 2\pi}\int{\cal K}_{\rm MM} M^{\dagger}_{kk'}
M_{pp'}\,\,.  }
It is now straightforward to recover 't Hooft equation by
evaluating the action of the operator $V_{\rm MM}$ on general
one-meson states of the form,
\eqn\unmeson
{|\phi, jj', P^{+} \ket = \sqrt{P^{+}} \int_{0}^{1} \d x\,\,\phi(x)
\,M^{\dagger} (xP^{+}, j; (1-x)P^{+}, j')\, |0\ket\,\,. }
This is a normalized one-meson state with
total light-cone momentum $P^{+}$, light-cone wave function
$\phi(x)$ and flavors $j$ and $j'$ for quark and antiquark,
respectively. The variable $x$ has the physical interpretation of the
fraction of light-cone momentum carried by the quark.

So, to leading order in $1/N$ the action of $P^{-}$ on the wave
function $\phi (x)$ is given by
\eqn\th
{P^{-} [\phi] = {1\over 2P^{+}} \left[\Bigl( {m^2_{j} \over x} +
{m^2_{j'} \over 1-x} \Bigr) \phi (x) - {g^2 N \over \pi}
\pvi_{0}^{1} \d y\,\, {\phi (y) - \phi (x) \over (x-y)^2} \right] +
{\rO}\left({1\over \sqrt{N}}\right)\,\,.  }
Here we have absorbed the mass shift ${\bar m}^2 = m^2 - g^2 N/\pi$
into the integral operator by means of the identity,
$$
\pvi_{0}^{1} {\d y\over (x-y)^2} = -{1\over x(1-x)}\,\,.  $$
As it was previously announced, the action of the
divergent piece on the hamiltonian $V_{\rm sing}$
on meson states is completely cancelled by
contributions from the $BB$ and $DD$ terms.
In eq.~\th\ we deal with the infrared divergences of the potential
energy by means of the principal value analytic regularization, which
is known to give the correct physical results for this
theory~\refs\bglcg.

There are corrections
to the bound-state equation coming from the subleading
``singlet" terms in the $MM$, $BB$, $DD$ and $BD$ kernels. Their total
effect amounts to a simple renormalization of the coupling constant
$g^2 \rightarrow {\bar g}^2 = g^2 (1-1/N^2)$. As a result, the
$1/N$-corrected bound-state equation, $\mu^2 = 2P^{+} P^{-}$, reads:
\eqn\thooft
{\mu_{n}^{2} \phi_n (x)=\Bigl({m_{j}^2\over x}+{m_{j'}^{2}\over
1-x} \Bigr) \phi_n (x) - {{\bar g}^2 N \over \pi} \pvi_{0}^{1}
\d y\,\, { \phi_n (y) - \phi_n (x) \over (x-y)^2}\,\,.  }
This is the standard 't Hooft equation with a definition of
constituent quark mass compatible with the appearance of a pion in the
chiral limit. Namely, as $m\rightarrow 0$ then $\phi = 1$ becomes a
solution with $\mu =0$.
Other, less ``physical", definitions of the quark mass are also
possible. For example, in the totally normal-ordered hamiltonian $m$
is {\it not} shifted by the coupling constant and  the large-$N$ chiral
symmetry point would be at $m^2= -g^2 N /\pi$.

We have seen that $1/N$ corrections to the 't Hooft equation coming
from the one-meson sector are relatively mild and amount
simply to a redefinition of the coupling. In fact, these corrections
are kinematical and do not occur for a U$(N)$ gauge group, since they
ultimately come from the traceless character of the
generators~\normT{{\rm b}}.
The remaining $1/N$ corrections are non-trivial and correspond to
meson exchange, as  will be discussed in sect.~6.

The power of the light-cone method in deriving bound-state equations
is not restricted to the meson sector. As an illustration we quote here
the bound-state equation for baryons in the large-$N$ limit~\refs\durgut,
which can be derived by acting with $P^{-}$ on baryon states of the form,
$$
|\CB\ket = \int_{\{k\}} \phi(k_1,...,k_N) {1\over \sqrt{N!}}\,\,
\epsilon^{\alpha_1 \dots \alpha_N} b^{\dagger}_{\alpha_1} (k_1)\dots
b^{\dagger}_{\alpha_N} (k_N) |0\ket \,\,. $$
Here we consider a single flavored ``proton" for simplicity. Acting
on this type of states only the $BB$ term in $V$ contributes to
leading order and we obtain,
$$\eqalign{
{\mu^2\over 2P^{+}}\,\, \phi(k_1,\dots,k_N)&=\sum_{i=1}^{N}
{{\bar m}^2\over 2k_i}\,\,
\phi(k_1,\dots,k_N) \cr
&- {g^2 N\over 2\pi} \sum_{i<j}^{N} {1\over N} \pvi
\d p\,\,{\phi(k_1,\dots,p,\dots,k_i+k_j-p,\dots,k_N)\over (p-k_i)^2}\,.}$$
This is a relativistic hamiltonian equation describing
pairwise interactions of order $1/N$ among $N$ quarks,
as expected on physical grounds.
The Hartree approximation proposed by Witten~\refs\witten\ in a
non-relativistic context can be implemented in this relativistic model
in explicit hamiltonian form. Finally, we note that an
effective hamiltonian to compute interactions could be derived along
the lines of sect.~2, with selection rules,
$$
{1\over N}\,H_{\rm eff}\sim H_{(1,1)} \CB^{\dagger}\CB
+{1\over 4}\,H_{(2,2)}\,
\CB^{\dagger}\CB^{\dagger}\CB\CB+{1\over\sqrt{N}}H_{{\CB}{\CM}}
({\CM}^{\dagger}\CB^{\dagger} \CB + \CB^{\dagger} \CB \CM )\,\,, $$
where the last term corresponds to meson emission by baryons.

\newsec{Effective meson hamiltonian}

In this section we construct the effective hamiltonian which
summarizes the $1/N$ expansion of two-dimensional QCD. The
derivation follows the general discussion in sect.~2, with the
meson creation operators given by:
\eqn\meson
{\CM^{\dagger}(n,jj',P^{+})=\sqrt{P^{+}}\int_{0}^{1} \d x\,\,\phi_n(x)
\,M^{\dagger} (xP^{+}, j; (1-x)P^{+}, j')\,\,,  }
where $\phi_n (x)$ is the solution of the bound-state equation \thooft.
These operators are normalized to leading order as,
\eqn\norm
{\bigl[\CM(n,jj', P^{+}), \CM^{\dagger}(m,ll', Q^{+})\bigr] =
\delta_{nm}\delta_{jl}\delta_{j'l'} \delta (P^{+} - Q^{+})\,\,. }
The commutator of $\CM$ and $\CM^{\dagger}$ contains some
subleading extra terms corresponding to the fact that the states
$\CM^{\dagger}|0\ket$ as defined in \meson\ are orthonormal
only in the large-$N$ limit. However, within $1/N$
perturbation theory we are interested in matrix elements of the
hamiltonian in the $N=\infty$ orthonormal basis. Therefore,
when computing
the nested commutators in \nested\ we use \norm, neglecting $1/N$
corrections to the normalization of the asymptotic states. In this way
we always work with the planar asymptotic states and ensure the
consistency of the method.

Using the basic commutators,
\eqn\bcom{\eqalign{
&\bigl[B_{kk'},M^{\dagger}_{pp'}\bigr]=\delta_{k'p}M^{\dagger}_{kp'}
\,\,,\cr
&\bigl[D_{kk'},M^{\dagger}_{pp'}\bigr]=\delta_{k'p'} M^{\dagger}_{pk}
\,\,,}}
the computation of the nested commutators in \nested\ is long but
entirely straightforward. We find the following structure:
\eqn\selrul
{\eqalign{P^{-}_{\rm eff}=H_{(1,1)}\CM^{\dagger}\CM &+ {1\over 2\sqrt{N}}
\left(H_{(1,2)}+ {1\over N} H'_{(1,2)}\right)
(\CM^{\dagger} \CM^{\dagger}\CM+\CM^{\dagger}\CM\CM ) \cr
\noalign{\vskip 0.2cm}
&+ {1\over 4N} \left(H_{(2,2)}+ {1\over N} H'_{(2,2)}\right)
\CM^{\dagger} \CM^{\dagger}\CM \CM\,\,, }}
where the primes refer to the ``singlet" subleading terms in the
kernels  \kernels.
An interesting feature of this hamiltonian is that it has only a
finite number of terms (cubic and quartic interaction). This is so
because in the light-cone gauge both single-meson states
\meson\ and the fundamental QCD hamiltonian \kin, \pot, \sing\
are polynomials in $M, M^\dagger, B, D$. From the properties of
commutators \norm\ and \bcom\ it then follows that the nested commutator
will eventually vanish. In the equal-time formalism meson states are
complicated ``coherent'' states in terms of $M^\dagger$ and it is not
a priori obvious that the effective hamiltonian would terminate.

After solving the  't Hooft equation, $H_{(1,1)}$ becomes the standard
kinetic energy for free mesons,
\eqn\freem
{P^{-}_{{\rm eff}, 0} = \sum_{n(jj')} \int_{0}^{\infty} \d P^{+}\,
{\mu_n^{2}(jj')\over 2P^{+}}\,\CM^{\dagger}(n,jj',P^{+})\,
\CM(n,jj', P^{+})\,\,.  }
The three-meson coupling picks contributions only from the
$M^{\dagger}B$  and $M^{\dagger} D$ terms,
\eqn\tressel
{{1\over \sqrt{N}} H_{(1,2)}+ {1\over N\sqrt{N}} H'_{(1,2)}  \sim
\bbra 0  \bigl| \CM\CM \bigl[ V_{\rm MB} + V_{\rm MD} ,
\CM^{\dagger}\bigr] \bigr|0\bket \,\,, }
while the four-meson vertex depends on the $BB$, $DD$ and $BD$ terms,
\eqn\cuatrosel
{{1\over N} H_{(2,2)} + {1\over N^2} H'_{(2,2)} \sim
\bbra 0\bigl| \CM\CM \bigl[\bigl[V_{\rm BB}+V_{\rm DD}+V_{\rm BD},
\CM^{\dagger}\bigr], \CM^{\dagger}\bigr] \bigr|0\bket\,\,. }

Before writing the answer for the vertices it is convenient to
introduce some notational devices.
Let us define the charge conjugation operator, $\CC$, which
interchanges quarks and antiquarks ($b^{\dagger}$ and $d^{\dagger}$
operators). Acting on wave functions and flavor indices we have,
\eqn\chargec
{\CC\bigl[f(j,j') \phi(x)\bigr] = f(j',j)\phi(1-x)\,\,. }
This operator simplifies the computations because it relates the
kernels in \pot\ as $\CC({\cal K}_{\rm MB})= -{\cal K}_{\rm MD}$ and
$\CC({\cal K}_{\rm BB})={\cal K}_{\rm DD}$. We also define the permutation
operators $\CP_L$ $(\CP_R)$ on initial (final) labels, with obvious
action on all quantum numbers and variables.
Finally, from all vertices we extract a factor of the form:
\eqn\ver
{{1\over N^{\rm power}}\,H_{({\rm in},{\rm out})}={g^2 N \over \pi}\,
{\delta(P_{\rm in}^{+} - P_{\rm out}^{+} ) \over
\prod\limits_{{\rm legs},l}
\sqrt{P_{l}^{+}}}\,\, [{\rm in} \rightarrow {\rm out}]\,\,.}

We classify vertices by the different flavor structures. For the
leading three-meson interactions \tressel\ we find two independent
flavor structures which we write as
\eqn\tv{\eqalign{
\bigl[L\rightarrow R_1 R_2\bigr] &={1\over 2\sqrt{N}}\,(1+ \CP_R)\,\,
\delta_{\ell r_1}\delta_{\ell' r_2'}\delta_{r_1' r_2}
\,\CT (L|R_1 R_2 ; \omega)\cr
\noalign{\vskip 0.3cm}
&+{1\over2N\sqrt{N}}\,(1+\CP_R)\,\delta_{\ell r_1}\delta_{\ell' r_1'}
\delta_{r_2 r_2'}\, \CT' (L|R_1 R_2; \omega)\,\,. }}
For example, the two flavor structures of the leading term in
eq.~\tv\ explicitly read:
$$ \delta_{\ell r_1}\delta_{\ell' r_2'}\delta_{r_1'  r_2}\,\,
\CT (L|R_1 R_2;\omega) +
\delta_{\ell r_2} \delta_{\ell'  r_1'}\delta_{r_2' r_1}\,\,
\CT(L| R_2 R_1;1-\omega)\,\,.$$
In the above formulae $\ell$, $L$ labels refer to initial
particles and $r$,
$R$ to final ones. Also $\omega \equiv P_{r_1}^{+} / P_{\rm out}^{+}$
denotes the light-cone momentum fraction carried by particle $R_1$.
$\CT$ is a convolution of wave functions which determines all
leading form factors~\refs\einh, and $\CT'$ is the subleading
triple-meson vertex.  All vertex functions are
explicitly listed in Appendix A.

Regarding the four-meson vertices \cuatrosel\ we separate
$BB+DD$ and  $BD$ contributions.  In the first case there are
two flavor structures coming from both the leading and the singlet
terms:
\eqn\fvbb{\eqalign{
\bigl[L_1 L_2 &\rightarrow R_1 R_2\bigr]_{{\rm B}^2+{\rm D}^2}
={1 \over 2N}\,
(1+\CP_R)\Bigl[\,\delta_{\ell_1 r_1}\delta_{\ell_1' r_2'}
\delta_{\ell_2 r_2} \delta_{\ell_2'  r_2'}
\,\CF_{{\rm B}^2+{\rm D}^2}
(L_1 L_2; z | R_1 R_2; \omega)\,\Bigr]  \cr
\noalign{\vskip 0.2cm}
&+{1\over 2N^2}\,(1+\CP_R)
\Bigl[\,\delta_{\ell_1 r_1}\delta_{\ell_1' r_1'}
\delta_{\ell_2 r_2} \delta_{\ell_2' r_2'}
\, \CF^{'}_{{\rm B}^2 + {\rm D}^2}
(L_1 L_2 ; z | R_1 R_2 ;\omega)\,\Bigr] \,\,,}}
where $z\equiv P_{\ell_1}^{+} / P_{\rm in}^{+}$ is the analog of $\omega$
for the initial state.
Finally, from the $BD$ terms we find four leading flavor structures
and two subleading ones:
\eqn\fvbd{\eqalign{
\bigl[L_1 L_2 &\rightarrow R_1 R_2\bigr]_{{\rm BD}}
={1\over 4N}\,(1+ \CP_L+ \CP_R+ \CC)\Bigl[
\delta_{\ell_1 r_1}\delta_{\ell_1'  \ell_2} \delta_{\ell_2' r_2'}
\delta_{r_1'  r_2}
\,\CF_{{\rm BD}}(L_1 L_2;z|R_1 R_2;\omega)\Bigr]\cr
\noalign{\vskip 0.2cm}
&+{1\over 2N^2}\,(1+\CP_R)\Bigl[\,\delta_{\ell_1 r_1}
\delta_{\ell_2  r_2} \delta_{\ell_2' r_2'}\delta_{\ell_1'  r_1'}
\,\CF^{'}_{\rm BD}(L_1 L_2; z|R_1 R_2; \omega) \,\Bigr]\,\,.}}

\ifigc\legend{Graphical representation of vertices
\tv, \fvbb\ and \fvbd. }
{\epsfxsize3.50in\epsfbox{legend.eps}}

\noindent
We represent vertices \tv, \fvbb\ and \fvbd\ graphically as
shown in figs.~\legend{a}, \legend{b} and \legend{c},
respectively.

With the explicit expressions for the vertex functions, $\CT$ and $\CF$,
listed in Appendix A, we have completed the description of the effective
meson hamiltonian of two-dimensional QCD, which may now be used to
compute systematic $1/N$ corrections to meson scattering amplitudes
and spectrum.

The solutions of 't Hooft equation are not known analytically and
we can only evaluate our vertex functions numerically.
To solve 't Hooft equation we used the
Multhopp's method to transform it
into an infinite system of algebraic equations which is then
solved approximately by truncation. This method was first applied
to eq.~\thooft\  in ref.~\refs\numer, and later used in
refs.~\refs{\gmm,\jaffe}. (For a nice summary of the method
which we closely follow, see appendix of ref.~\refs\jaffe.)
For moderate mass of the light quark results are
independent of truncation already with several hundreds of
equations. Near the chiral limit
$(m_q\to 0)$, the convergence becomes quite slow and the
discretized light-cone method may be a better option.
Some examples of $\CT$ functions can be seen in
fig.~5 in sect.~8.

Since we have a light-cone hamiltonian we must use
light-cone perturbation theory, which is in fact a time-ordered
perturbation scheme (see, c.f., ref.~\refs\paulib). It is very similar to
old-fashioned perturbation theory, although
it is much simpler due to the absence of vacuum subgraphs (there are
no pure positive/negative frequency terms in the light-cone
hamiltonian). It is defined through the Born expansion of the complete
propagator between light-cone wave functions:
\eqn\pertth
{\eqalign{\Bigl\langle\,{\rm out}\,\Big\vert
{1\over E - P^{-}_{\rm eff} + i 0 }\Big\vert\,{\rm in}\,\Bigr\rangle
&=\Bigl\langle\,{\rm out}\,\Big\vert {1\over E - P_{0}^{-}+i 0}
\Big\vert\,{\rm in}\,\Bigr\rangle   \cr
\noalign{\vskip 0.2cm}
&+\Bigl\langle\,{\rm out}\,\Big\vert\,
{1\over E - P_{0}^{-} +i 0}\, P_{I}^{-}\,{1\over E - P_{0}^{-}+i 0}+\dots
\Big\vert\,{\rm in}\,\Bigr\rangle\,\,. }}
Here $E\equiv i\partial / \partial x^+$ is the light-cone energy
operator, and we split the effective hamiltonian into free and
interaction parts
$$
P_{\rm eff}^{-} = P_{0}^{-} + P_{I}^{-}\,\,, $$
with $P_{0}^{-}$  given by \freem.
We evaluate \pertth\ by inserting the spectral decomposition of
unity between any two interactions:
$$
1=\sum_{\{n(jj')\}}\int \{\d P^{+} \}\,\,{\CM^{\dagger}_1\dots
\CM^{\dagger}_K|0\ket \bra 0 | \CM_1 \dots\CM_K \over K!} \,\,.$$

This algorithm determines the corresponding graphical rules, which
have the same symmetry factors as the Feynman rules once the factor
$1/p!q!$ has been extracted from each vertex as in eq.~\effham.  One could
also recover this perturbative expansion from covariant perturbation
theory in the fundamental quark/gluon variables. After going to the
light-cone gauge, the integrals over intermediate $k^{-}$ components
of momenta can be evaluated by residue calculus. For each pole one
finds the corresponding time-ordered process of light-cone
perturbation theory. Within Feynman's $i\epsilon$ prescription
and 't Hooft's principal value prescription one can explicitly
show the equivalence diagram by diagram~\refs\pert. The main advantage of our
method is that the gluons are integrated out from the very beginning
and there is no need to consider Bethe-Salpeter techniques.

\newsec{$1/N$ corrections to the spectrum}

With the effective hamiltonian \selrul\ at hand we may compute the
mixing mass matrix for mesons in light-cone perturbation theory by the
usual procedure of summing the one-particle irreducible (1PI)
geometric series for the propagator:
$$
\left({1\over p^2 - \mu^2}\right)_{n,n'} = \left( {1\over p^2 -
\mu^{(0) 2} - \Sigma (p^2)}\right)_{n,n'} $$
and we obtain the mixing matrix:
\eqn\mixing
{\mu_{nn'}^{2} = \mu_{n}^{(0) 2} \delta_{nn'}+\Sigma_{nn'} (p^2)\,\,, }
where $\mu_{n}^{(0) 2}$ stands for the large-$N$ spectrum of
eq.~\thooft. The leading $1/N$ contribution to the self-energy matrix
$\Sigma_{nn'}$ comes from bubble diagrams with three-meson vertices
\tv. Let us consider, as an example, a two-flavor model with heavy
$(Q)$ and light $(q)$ quarks. We can then form two-dimensional analogs
of pions ($q{\bar q}$), $B$ mesons ($Q{\bar q}$),
${\bar B}$ mesons ($q {\bar Q}$) and $\eta_b$ mesons ($Q{\bar Q}$).

\ifigc\BBmixing{Diagrams contributing to $B_n$--$B_{n'}$ mixing.}
{\epsfxsize4.5in\epsfbox{BBmixing.eps}}

For example, to leading order there is no $B$--$\pi$ or
$B$--$\eta$ mixing.  The leading contributions to
$B_n$--$B_{n'}$ mixing are given by the diagrams in fig.~\BBmixing.
For the physical values of the masses, diagrams in fig.~\BBmixing{b}
are strongly supressed compared to those in fig.~\BBmixing{a}.
We quote here the result for $\pi, B$ exchange -- diagram \BBmixing{a}:
\eqn\self
{\Sigma_{nn'} (p^2)\bigg\vert_{B,\pi\,\,{\rm channel}}
= -{1\over N} \left( {g^2 N\over \pi}\right)^2
{1\over p^2} \sum_{k,m} \int_{0}^{1} \d z\, {\CT(B_n |B_k \pi_m ; z)
\CT(B_{n'} | B_k \pi_m ; z) \over (z-z_{+})(z-z_{-})}\,\,,  }
where
$$
z_{\pm} = {1\over 2p^2} \left( p^2 + \mu_{+}\mu_{-} \pm \sqrt{(p^2 -
\mu_{+}^{2})(p^2 - \mu_{-}^{2})}\right)\,\,, $$
and $\mu_{\pm} \equiv \mu_{k}^{(0)} \pm \mu_{m}^{(0)} $, with the
standard $i\epsilon$ prescription implicit in \self.
This formula clearly exhibits the threshold structure.
In particular, in a low-energy
approximation we may simplify expression \self\ by setting the
vertex functions equal to the on-shell values $\CT_{nkm}$, and we get
the standard form of the bubble diagram:
$$
\Sigma_{nn'} (p^2) \simeq -{1\over N} \left({g^2 N \over \pi}\right)^2
\sum_{k,m} {\CT_{nkm} \CT_{n' km} \over \sqrt{(p^2 -\mu_{+}^2)(p^2 -
\mu_{-}^2)}}\, {\rm log} \left[\,{\sqrt{\mu_{+}^2 - p^2} + \sqrt{\mu_{-}^2
- p^2} \over \sqrt{\mu_{+}^2 - p^2} - \sqrt{\mu_{-}^2 - p^2}}
\,\,\right]^2\,\,, $$
where we recognize the usual threshold function.

We have calculated numerically several elements of the $\Sigma$-matrix
as given by
formula \self\ for three different values of the light quark mass:
$m_q^2 = 0.3136, 0.1 , 0.01$ , and for the heavy quark mass
$m_Q^2 = 2000$. In the units ${\bar g}^2 N / \pi = 1$, the large-$N$ value of
the $B$ meson ground state mass is found to be
$\mu_B^{(0) 2} = 2121, 2101, 2089$, respectively.
The on-shell values of $\Sigma_{00}$ for $N=3$ are
found to be: $-52, -318, -1560$.
(Note the negative sign, typical of second order perturbation theory
about the ground state.) We have
checked that the contributions with larger values of the excitation
numbers, $k$ and $m$, in eq.~\self\ quickly become very small. We note,
however, that the largest contributions to $\Sigma_{00}$ come from
diagrams involving odd $B$-resonances in the intermediate state, and
similar results are obtained for $\Sigma_{11}$ and $\Sigma_{22}$. The
vertex functions with odd $B$-resonances in the final state become
very large as we approach the chiral limit for the light quark mass.
In addition, the bubble diagram for the diagonal elements of the
self-energy matrix is anomalously large, because the $\cal T$
functions enter squared, and no destructive interference between the
vertices occurs. This symmetry enhancement is not present for the
off-diagonal elements of the self-energy matrix. For example, for the mixing
between the first $B$-resonance and the fundamental state we find
$\Sigma_{01} = 7, 116, 760$ for the different light quark masses.
We see that the mixings are much smaller than the diagonal level
shifts, but still, they seem to blow up as we approach the chiral limit
$m_q \rightarrow 0$. These uncontrolable large corrections for light
quark mass do not mean that the $1/N$ expansion is sick, but rather
that the chiral limit does not survive $1/N$ corrections. In fact, the
renormalization procedure that fixes the quark masses must be repeated
at any order in the perturbative expansion, and the large values of
the self-energy matrix translate into large renormalizations of the
quark masses, but still small overall corrections to the physical
spectrum.

To see how this works, let us denote by $\sigma_n (p^2)$ the
eigenvalues of the self-energy matrix. These quantities depend on
$p^2$ and also on the dimensionless parameters of the
the 't Hooft equation:
\eqn\param
{x_q={\pi m_{q}^2\over{\bar g}^2 N}\,\,,\qquad\quad x_Q =
{\pi m_{Q}^2 \over {\bar g}^2 N}\,\,.  }
To any order in $1/N$ the quark masses are determined by fitting the
mass of the fundamental states $\pi_0$ and $B_0$ to the ``experimental"
pion and $B$ meson masses:
\eqn\spefc
{\mu_{0}^{(0) 2} (x_q , x_Q) + \sigma_0 (x_q , x_Q , \mu_{0}^2 ) =
\mu_{0}^2\,\,.}
Taking  $\mu_{0}$ to be the mass of pion or $B$ meson we get two equations
that fix $x_q$ and $x_Q$. The rest of the spectrum $\mu_{n}^2$ is then
obtained from the following implicit equation:
\eqn\spec
{\mu_{n}^{(0) 2} (x_q , x_Q) + \sigma_n (x_q , x_Q  , \mu_{n}^2) =
\mu_{n}^2 \,\,. }

We can illustrate the renormalization of the quark masses in a simple
example.  For a qualitative discussion,
let us neglect the heavy quark mass running and fix $x_Q = 2000$.
We will consider the contribution to the light quark renormalization
coming from the $B$-resonances only, and furthermore neglect the
off-diagonal mixings.  We define the mass of the fundamental $B$
meson to be:
$$ 2101 = \mu_{0}^{(0)2} (x_q = 0.1)\,\,.  $$
This equation means that the mass of the light quark at $N=\infty$ is
taken to be $x_q = 0.1$. The $1/N$-corrected value of $x_q$ is
obtained by solving the equation \spefc
$$ 2101 = \mu_{0}^{(0)2} (x_q) + \sigma_0 (x_q , p^2 = 2101)\,\,. $$
A simple estimate follows from linear interpolation between two points:
at $x_q = 0.1$,  $\sigma_0 (p^2 = 2101) \sim \Sigma_{00} (p^2 = 2101)
\sim -320$, and at $x_q = 0.3136$, $\sigma_0 (p^2 = 2101) \sim -52$.
This results in an estimate of the renormalized quark mass of $x_q
\simeq 0.34$. We obtain a larger mass, for which the self-energy
corrections are again small. For this value of the quark masses, the
typical shifts found from equation (6.5) for $N=3$  are,
$$ {\delta \mu_n^2 \over \mu_n^2} \sim 2.5 \% \,\,. $$
So, taking into account that the self-energy diagram is
anomalously large, as explained before, we see that the typical $1/N$
corrections are small, and the apparent singularity for very light
quarks
simply means that the interacting theory restores the chiral
symmetry, if broken in the free $N=\infty$ theory. In general, the
fact that the self-energy corrections are large and negative as $m_q
\rightarrow 0$, implies that the chiral symmetry renormalization point
$m_q =0$ is unstable under $1/N$ corrections, and the whole picture
agrees with Coleman's theorem, which forbids spontaneous symmetry
breaking in two dimensions.

As a final comment, we point out that $\sigma_n$ also induces a wave
function renormalization,
\eqn\wren
{Z_n = {1\over 1- {\partial\sigma_n \over \partial p^2} (p^2 =
\mu_{n}^2)}\,\,,  }
which in turn amounts to extra $1/N$ corrections to the vertices, to be
added to the trivial $1/N^2$ correction coming from the
renormalization of $g\rightarrow {\bar g}$:
\eqn\verren
{\CT(L|R_1 R_2)\rightarrow\sqrt{Z_L Z_{R_1}Z_{R_2}}\,\,
\CT(L|R_1 R_2)\,\,,}
and similarly for the $\CF$ functions.

\newsec{$1/N$ corrections to form factors}

The planar limit of the electromagnetic form factors of
$(1+1)$-dimensional QCD was derived in ref.~\refs\einh. Recently,
in ref.~\refs\gm, the analysis was extended
to the case of flavor changing currents, in order to simulate
semileptonic heavy meson decays in two dimensions.
According to ref.~\refs\gm\ the form factors for semileptonic decays of $B$
mesons are single-pole dominated in the combined limits: $N
\rightarrow\infty$, $M_b \rightarrow\infty$ and $m_{\pi} \rightarrow
0$ in the four-dimensional theory. The argument goes schematically as
follows: one first writes the form factor as a superposition of pole
terms, according to the spectral decomposition at $N=\infty$ (narrow
width approximation),
$$
F(q^2) \sim \sum_n {C_n\over q^2 -\mu_{n}^2 }\,\,, $$
and then uses the heavy quark limit to prove the vanishing of all
$C$'s but one, by means of chiral Ward identities.

This argument was tested in ref.~\refs\gmm\ in planar two-dimensional
QCD. It turns out that  the heavy quark mass limit is not neccesary
in two dimensions due to the constrained kinematics, and the single
pole dominance is easily obtained in the chiral limit for the pions.
The nature of the $1/N$ corrections to this behavior can be easily
studied using the formalism we have developed here.
As an example,
we will again consider a two-flavor model as in sect.~6.
Since two-dimensional vector and axial currents are dual of each
other, we only need to consider the coupling of mesons to
the flavor-changing vector current:
$$
J_{Qq}^{\mu} = {\bar \psi}_{Q} \gamma^{\mu} \psi_{q}\,\,.  $$
In order to derive effective meson--current vertices to be used in the
$1/N$ perturbation theory, we will couple the quark current to a
background vector field in a minimal fashion:
\eqn\eqc
{P_{Qq}^{-} = e\int\limits_{x^+ = {\rm const.}}
\Bigl( J_{Qq}^{\mu} W_{\mu} \,\,+ {\rm h.c.}\Bigr)\,\,. }
After solving for the physical fermion fields we have,
\eqn\qcc
{P_{Qq}^{-} = e\sqrt{2} \int\limits_{x^+ ={\rm const.}}\left(
:\psi_{Q}^{\dagger} \psi_{q} : W_+ +{m_q m_Q \over 2}
:\Bigl({1\over i\partial_{-} }\psi_Q \Bigr)^{\dagger}
\Bigl({1\over i\partial_{-} }\psi_q \Bigr) :
W_{-} + {\rm h.c.} \right)\,\,.   }
Defining a Fock space for the vector bosons,
\eqn\fs{\eqalign{
& W^{+}=( W^{-})^{\dagger}={1\over\sqrt{2\pi}}\int_{0}^{\infty}
{\d k^+\over\sqrt{2 k^+}}\left( a(k^+)\,\e{-ik^+ x^-} + a^{\dagger}(k^+)
\,\e{ik^+ x^-} \right)\,\,,  \cr
\noalign{\vskip 0.2cm}
& [a(k^+), a^{\dagger}(k'^+)] = \delta(k^+-k'^+)\,\,,   }  }
we derive a meson-vector boson effective hamiltonian along
the lines of sect.~2 with the result:
\eqn\ehff
{\eqalign{P^{-}_{Qq,{\rm eff}}&=\sqrt{N}\left( H_{(M\rightarrow W)}\,
a^{\dagger} \CM  + H_{(W\rightarrow M)}\, \CM^{\dagger} a \right) \cr
&+ H_{(M\rightarrow M+W)} \,a^{\dagger} \CM^{\dagger} \CM +
H_{(M+W\rightarrow M)}\, \CM^{\dagger} \CM a \,\,.} }
An interesting feature of this hamiltonian is the absence of direct
$W\rightarrow M+M$ or $M+M\rightarrow W$ transitions. This fact will
be of some significance in the sequel. The effective vertices in
\ehff\ may be computed from the general formula \nested, with the
result:

\noindent{$\bullet$} $B-W^{\pm}$ mixing:
\eqn\bwmixing
{\eqalign{&  H^+ \bigl[B_n (P^+) \rightarrow W^+ (q^+)\bigr]
= {e\over 2\pi}\delta (P^+ - q^+) {1 \over \sqrt{P^+}}
\int_{0}^{1} \d x\,\phi_{B_n} (x)\,\,, \cr
\noalign{\vskip 0.2cm}
& H^- \bigl[B_n (P^+) \rightarrow W^- (q^+)\bigr]
= -{e\over 2\pi} \delta (P^+ -q^+) {m_q m_Q \over 2 q^+ \sqrt{P^+}}
\int_{0}^{1} \d x \,{\phi_{B_n} (x) \over x(1-x)}\,\,,   }  }

\noindent{$\bullet$} $W^{\pm}$ emission by $Q$
($z\equiv q^+ / P^{+}_{B}$):
\eqn\wemission
{\eqalign{ H^+ \bigl[B_n (P^{+}_{B})\rightarrow \pi_m (P^{+}_{\pi})
&+W^+(q^+)\bigr]=
{e\over 2\pi}\,\delta (P^{+}_{B} - q^+ - P^{+}_{\pi})
\sqrt{P^{+}_{\pi} \over P^{+}_{B}} {1\over q^+} \cr
&\times \int_{0}^{1} \d x\, \phi_{B_n} (z +(1-z)x)\phi_{\pi_m}(x)\,\,,\cr
\noalign{\vskip 0.2cm}
H^- \bigl[B_n (P^{+}_{B}) \rightarrow\pi_m (P^{+}_{\pi})
&+W^-(q^+)\bigr]={e\over 2\pi}
\delta (P^{+}_{B} -q^+ -P^{+}_{\pi})
\sqrt{P^{+}_{\pi} \over P^{+}_{B}}{1\over q^+} \cr
\noalign{\vskip 0.2cm}
&\times {m_q m_Q \over 2  P^{+}_{\pi} P^{+}_{B}}
\int_{0}^{1} \d x\,\, {\phi_{B_n} [z +(1-z)x] \phi_{\pi_m} (x) \over
(1-x)[1-z-(1-z)x] } \,\,. }  }
Analogous expressions hold for $W^{\pm}$ emission by ${\bar q}$, with
the following modifications of \wemission: there is an overall minus
sign, $\pi_m$ is substituted by $\eta_m$ in the final state, and $z
+(1-z)x \rightarrow (1-z)x$ inside the integrals.

Given the complete effective hamiltonian, including \ehff,  we are
in position to compute form factors. Let us consider the time-like
form factor defined as the light-cone 1PI amplitude for
$W^{\pm}$ decay into on-shell $B$ and $\pi$ mesons:
\eqn\ff
{F^{\pm} (q^2) = C \Bigl\langle B(\omega q^+); \pi((1-\omega)q^+)
\Big\vert (q^- - P_{0}^{-}) {1\over q^- - P^{-}_{\rm eff} +i0} (q^- -
P_{0}^{-})\Big\vert W^{\pm} (q^+)\Bigr\rangle_{1{\rm PI}}\,\,.  }
In this formula the factors $(q^- -P^{-}_{0})$ amputate external
propagators, and the constant
$$
C= {4\pi \over e} q^+ \sqrt{P_{B}^{+}P_{\pi}^{+}} $$
defines $F^{\pm}$ compatible with the conventions in ref.~\refs\gmm.
Also,
$\omega \equiv P_{B}^{+} /q^+$ is the light-cone momentum fraction
carried by the $B$ meson. It becomes a function of $q^2$ once
conservation of light-cone energy is enforced
for on-shell $B$ and $\pi$ mesons,
\eqn\onshell
{q^2 = {\mu_{B}^2 \over \omega} + {\mu_{\pi}^2 \over 1-\omega}\,\,.  }

\ifigc\formfactor{Form factor diagram for the decay
$W^{\pm}\to B + \pi$.}
{\epsfxsize2.2in\epsfbox{ffactor.eps}}

The $1/N$ expansion of the form factor is generated by
expanding the propagator in \ff.
The first term (free propagation) vanishes because of
the kinematics, and the second term (one intermediate interaction) also
vanishes because there are no couplings for direct decay of $W^{\pm}$
into two mesons in our effective hamiltonian \ehff. So, the
${\rO}(1)$ contribution involves at least two intermediate interactions,
and we can separate it from the $1/N$ corrections
as (see fig.~\formfactor):
\eqn\split
{\eqalign{ F^{\pm}& (q^2) =  \sum_{M_1,M_2} \Bigl\langle B\pi
\Big\vert (q^- - P^{-}_{0} ) {1\over q^- - P^{-}_{\rm eff} +i0} (q^- -
P^{-}_{0}) \Big\vert \,M_1 M_2\Bigr\rangle_{1{\rm PI}} \cr
\noalign{\vskip 0.2cm}
& \times C \Bigl\langle\, M_1 M_2\Big\vert {1\over q^- - P^{-}_{0}
+i0}{H(2\leftarrow 1) \over 2\sqrt{N}}
 {1\over q^- - P^{-}_{0} +i0}\sqrt{N} H_{(B_n
\leftarrow W^{\pm})} \Big\vert W^{\pm} \Bigr\rangle_{1{\rm PI}}\,\,, }}
where $H(2\leftarrow 1)$ stands for $H_{(2,1)} + H'_{(2,1)} /N$.
The intermediate state of two mesons is either a $B_{\ell}$, $\pi_m$
state or a $B_{\ell}$, $\eta_m$ state, and the corresponding propagator
and measure contribute a factor:
$$
q^{+} \int_{0}^{1} \d z\,\, {2q^{+} \over q^2 - {\mu_{\ell}^2 \over z} -
{\mu_{m}^{2} \over 1-z} }\,\,,$$
where $\mu_{\ell}$ is the mass of $B_{\ell}$ and $\mu_m$ that of
$\pi_m$ or $\eta_m$. Also, $z\equiv P_{B_{\ell}}^{+} / q^+$ is the
light-cone momentum fraction of the $B_{\ell}$ resonance, and the
extra factor $q^+$  is the Jacobian from the intermediate
light-cone momenta to the momentum fraction $z$. The remaining
propagator in formula \split\ carries a $B_n$ resonance of mass
$\mu_n$, created out of the vacuum by the current, and it contributes
a factor $ 2q^+ /(q^2 - \mu_n^2)$.

Now, putting everything together, in terms of the meson decay
constant, $f_n \equiv \int \d x\,\phi_n (x)$,  we can write:
\eqn\dosf
{\eqalign{& F^+ (q^2)=2q^+\,{g^2 N\over\pi}\,\sum_n {f_n T_n (q^2)
\over q^2 - \mu_n^2 }\,\,,  \cr
\noalign{\vskip 0.2cm}
& F^- (q^2)=-{1\over q^+}\,{g^2 N\over\pi}\,\sum_n {(-)^n f_n \mu_n^2
T_n (q^2) \over q^2 - \mu_n^2 }\,\,.  } }
In the second relation we have made use of the so-called parity
relation,
$$
m_q m_Q\int_{0}^{1}\d x\,{\phi_n(x)\over x(1-x)}=(-)^n \mu_n^2 f_n\,\,.
$$
which was proved in ref.~\refs\gross. The function $T_n (q^2)$ contains
all $1/N$ corrections and is of the form:
\eqn\tn
{T_n (q^2) = \CT (B_n | B\,\pi;
\omega) + {1\over N} \CT' (B_n | B\,\pi;\omega)+\Delta T_n(q^2)\,\,.}

\ifigc\corrections{Diagrams contributing to various four-point
Greens's functions.}
{\epsfxsize4.20in\epsfbox{fourpoint.eps}}

\noindent
Here we separated the leading term from the first correction,
given by the singlet three-meson coupling, and $\Delta T_n$
which includes meson exchange:
\eqn\deltatn
{\eqalign{ \Delta T_n (q^2)& = \sum_{\ell,m} \int_{0}^{1} \d z\left[ \CT
(B_n | B_{\ell}\, \pi_m ; z) + {1\over N} \CT' (B_n | B_{\ell}\, \pi_m ;
z) \right]  \cr
&\times \sqrt{\omega (1-\omega) \over z(1-z)} {2 (q^+)^2 \over q^2 -
{\mu_{\ell}^2 \over z} - {\mu_m^2 \over 1-z} }\biggl[ G(B_{\ell}\,\pi_m
;z|B\,\pi;
\omega) + G(\pi_m\, B_{\ell} ; 1-z| B
\,\pi; \omega)\biggr]  \cr
&+ \sum_{\ell,m} \int_{0}^{1} \d z \left[ \CT(B_n|\eta_m\, B_{\ell}; 1-z)
+ {1\over N} \CT' (B_n | B_{\ell}\, \eta_m ; z) \right]   \cr
&\times \sqrt{\omega (1-\omega)\over z(1-z)} {2 (q^+)^2 \over q^2
-{\mu_{\ell}^2 \over z} - {\mu_m^2 \over 1-z}} \biggl[ G(B_{\ell}\,\eta_m;
z|B\,\pi;\omega)+G(\eta_m\,B_{\ell};1-z|B\,\pi;\omega)\biggr]\,\,. }}
$G(2 {\rm meson}| B\,\pi ;\omega)$ represents the amputated Green's
function for the $(2\,{\rm meson} \rightarrow B\,\pi)$ transition in
formula \split. To leading order ($1/N$), there are nine diagrams
contributing to the various four-point Green's functions $G$. They
are listed in fig.~\corrections\
with the correspondig combinatorial factors. We
have either irreducible four-meson vertices for direct processes like
$(B_{\ell},\pi_m)\rightarrow (B,\pi)$, or two sucessive three-meson
interactions coming with two different
time orderings, like $ (B_{\ell},\pi_m)
\rightarrow (B_{\ell},\pi_k,\pi) \rightarrow (B,\pi)$ and
$(B_{\ell},\pi_m)\rightarrow (B,\pi_k,\pi_m)\rightarrow (B,\pi)$. For
example, the total ${\rO}(1/N)$ contribution to $\Delta T_n$ due to
$B,\eta$ exchange is given by the
diagrams \corrections{c}, \corrections{h} and \corrections{i}.
The remaining six diagrams correspond to $B,\pi$ exchange.
Because of the propagator supression  diagrams
\corrections{f} and \corrections{g} are smaller than
diagrams \corrections{a}, \corrections{b},
\corrections{d} and \corrections{e} which read:
\eqn\diag{\eqalign{
\Delta &T_n (q^2)_{B,\pi\, {\rm channel}}=
-{1\over N}{g^2 N\over \pi} \sum_{\ell,m} \int_{0}^{1} \d z\,
{\CT (B_n|B_{\ell}\,\pi_m ;z) \over q^2 (z-z_+)(z-z_-)} \cr
\noalign{\vskip 0.2cm}
&\times\biggl\{\,
\CF_{{\rm B}^2+{\rm D}^2}(B_{\ell}\,\pi_m;z|B\,\pi;\omega)
+{1\over 2}\,\CF_{\rm BD}(B_{\ell}\,\pi_m;z|B\,\pi;\omega) \cr
\noalign{\vskip 0.2cm}
&-2\,{g^2 N\over \pi}\sum_{{\scriptstyle k \atop \scriptstyle k+m={\rm
odd}}}
\biggl[\,\theta(\omega -z)\,\omega z\,
{\CT(B| B_{\ell}\,\pi_m; {z\over \omega})
\CT(\pi_m|\pi_k\,\pi; {\omega -z \over 1-z})
\over \mu_{B}^2 (z-\omega z_{+}')(z-\omega z_{-}')} \cr
\noalign{\vskip 0.2cm}
&+\theta(z-\omega)\,(1-\omega)(1-z) \,
{\CT(B_{\ell}| B\, \pi_m; {\omega \over z})
\CT(\pi|\pi_k \,\pi_m; {z-\omega \over 1-\omega})
\over\mu_{\pi}^2 (z-(1-\omega)z_{+}'')(z-(1-\omega)z_{-}'') }\,\biggr]
\,\biggr\} + {\rO}\Bigl({1\over N^2}\Bigr)\,\,, } }
where $z_{\pm} $ variables are defined in terms of the function
\eqn\zs{
Z_{ab}^{\pm} (p^2)\equiv{1\over 2p^2} \left(
p^2 +\mu_+\mu_-\pm\sqrt{(p^2 -\mu_{+}^2)(p^2 -\mu_{-}^2)}\right)\,\,,
\qquad\mu_{\pm} \equiv \mu_a \pm \mu_b \,\,, }
by the equations:
$$
z_{\pm} = Z_{\ell m}^{\pm} (q^2)\,,\qquad
z_{\pm}' = Z_{m k}^{\pm} (\mu_{B}^2) \,,\qquad
z_{\pm}'' = Z_{\ell k}^{\pm} (\mu_{\pi}^2)\,\,.  $$

We see how light-cone perturbation theory makes explicit the spectral
decomposition (unitarity). Looking at the cuts induced by the
functions \zs\ we identify the thresholds for $B_n \rightarrow
B_{\ell},\pi_m$ in all diagrams [from the common factor in \diag ].
The thresholds for $\pi_m, B_{\ell} \rightarrow B$ and
$\pi_k, \pi_m \rightarrow \pi$ can also be identified in the last two
diagrams, as functions of $\mu_{B}^2$ and $\mu_{\pi}^2$, respectively.
As usual, the pseudo-thresholds at $\mu_-$ should fall on unphysical
sheets.

In order to estimate the size of $1/N$ corrections to the form factors,
we evaluated numerically some contributions to $\Delta T_n(q^2)$
for $B,\pi$ channel given by expression \diag. We performed
calculations for $n=2$  (which corresponds to the first state above
threshold), and at the on-shell value of $q^2=\mu_{B_2}^2$.
The bare quark masses we used are: $m_q^2=0.3136$ and $m_Q^2=2000$.
The leading order contribution to the form factor in this case
is $\CT(B_2,B\pi;\omega_{\rm on-shell})\approx 37$.
We estimated a number of terms in the sum in eq.~\diag\ and found that
the largest contributions from the $\CT\times\CF$ terms are
$\approx 0.5-0.1$, while those of $\CT\times\CT\times\CT$
are $\approx 5\times 10^{-3}$. Also, these values decrease for
higher resonances in the intermediate states. We also found that
not all terms we calculated have the same sign which makes bounding
the total contribution a little difficult. However, it is safe
to say that the correction is not larger than $1-2 \%$.
We also mention that the $1/N$ correction coming from the
subleading three-meson coupling
${1\over N}\,\CT'(B_2,B\pi;\omega_{\rm on-shell})\approx 0.8$
(for $N=3$).
Finally, we expect these corrections to become much larger
as we approach the chiral limit (as was the case with corrections
to the spectrum discussed  in sect.~6). When computing form factors,
however, one should use the renormalized values of the quark masses,
and our previous analysis suggests that even if we start with
a light quark mass close to the chiral limit, $1/N$ corrections
will renormalize it to $\approx 0.3-0.4$.
This explains our choice of parameters mentioned above.
Although it might be interesting to explore these questions
in more detail, as well as to estimate
$\Delta T_n(q^2)$ for different values of $q^2$
(as is shown for the leading contribution in Fig.~5),
it is beyond our computational resources at this time
to perform such calculations.

The $1/N$ expansion we have derived for $F^{\pm} (q^2)$ is all we need to get
both vector and axial current form factors.  It is convenient to define the
combinations $f_{\pm} (q^2)$ through:
\eqn\dual
{\eqalign{
&q_{\mu}F^{\mu}(q^2)=(\mu_B^2 -\mu_{\pi}^2)f_+(q^2)+q^2 f_-(q^2)\,\,, \cr
\noalign{\vskip 0.2cm}
&\epsilon_{\mu\nu} q^{\mu}F^{\nu} (q^2) = 2 \epsilon_{\mu\nu}
p_{B}^{\mu} p_{\pi}^{\nu} f_+ (q^2)\,\,.  } }
In solving for $f_{\pm}$ from \dual\ and \dosf\ one encounters some
non-trivial steps. Namely, non-pole terms vanish because of the
identity\footnote{$^{*}$}{The same identity was found to leading order
in ref.~\refs\gmm.}
$$
\sum_{n=0}^{\infty}  f_n \, T_n (q^2) =0\,\,.  $$
This relation can be proved using the explicit form of
$\CT(B_n |2\,{\rm mesons})$ and $\CT'(B_n |2\,{\rm mesons})$ listed in
Appendix A, and the identity:
\eqn\ident
{\sum_n \, f_n \, \phi_n (x) = 1\,\,,}
which for $x\in [0,1]$ follows from the definition of $f_n$ and the
completeness of the wave functions \refs\einh.

Defining the generalized coupling $g_n$ as
\eqn\gene{\eqalign{
&g_n (q^2)={2\mu_n^2 \over {\mu_B^2 \over \omega} -q^2 \omega}\,\,
{g^2 N\over \pi}\, T_n (q^2)\,\,,\qquad\,{\rm for\,\,even\,} n\,\,,\cr
&g_n (q^2)=2\mu_n^2\,{g^2 N\over \pi}\, T_n (q^2)\,\,,
\quad\qquad\qquad{\rm for\,\,odd\,} n\,\,.  }}
the final result for $f_{\pm}$ reads:
\eqn\fin {\eqalign{
&f_+(q^2)=\sum_{{\rm even}\,n}\,\,{f_n g_n\over q^2-\mu_n^2}\,\,, \cr
\noalign{\vskip 0.2cm}
&f_- (q^2) = {1\over q^2} \,
\sum_{{\rm odd}\,n}\,\,{f_n g_n\over q^2-\mu_n^2}-{\mu_B^2 - \mu_{\pi}^{2}
\over q^2}\sum_{{\rm even}\,n}\,\,{f_n g_n \over q^2 - \mu_n^2}\,\,.  } }
So, we see that the $1/N$ expansion we derived for $T_n (q^2)$
contains  the corrections to $f_{\pm} (q^2)$. In particular,
keeping the leading term in \tn\ and recalling the expression for
the three-point vertex $\CT$ (Appendix A)
we make contact with the result of Grinstein and Mende~\refs\gmm.

\ifig\rolling{Vertex function $\CT(B_2|B_0,\pi_0;\omega)$
for the heavy quark mass $m_Q^2=2000$ and two different values of
the light quark mass: (a) $m_q^2=0.3136$, (b) $m_q^2=0.01$.
Dots mark the values of the on-shell couplings.}
{\epsfxsize3.90in\epsfbox{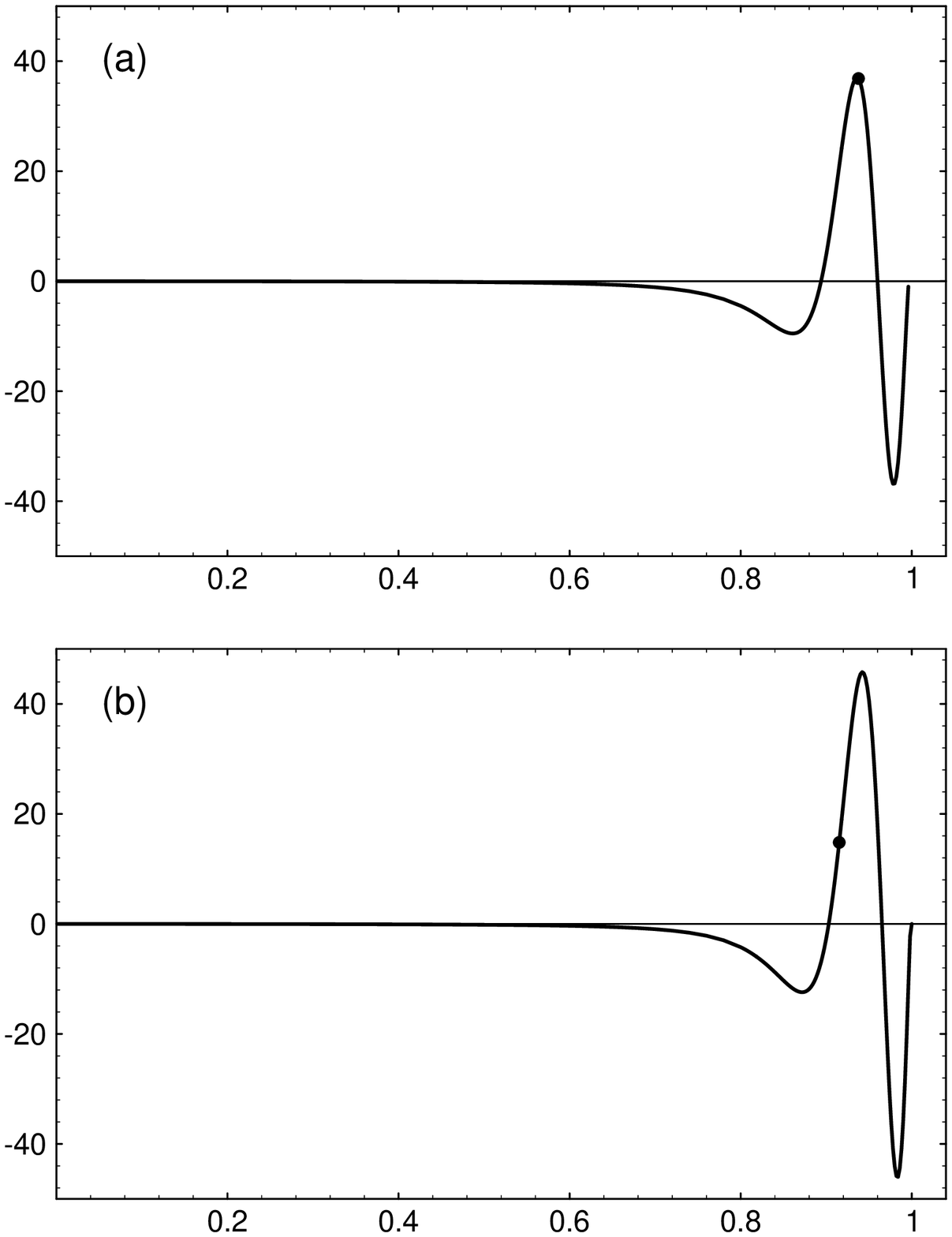}}

The couplings $g_n$ for even and odd $n$
(odd and even parity, respectively)
are related to the couplings appearing in a low energy meson effective
lagrangian. In the notation of ref.~\refs\gmm\ we have the tree level
structures:
$$
{\cal L}_{\rm int} = \sum_{abc} {1\over\sqrt{N}}\, {\hat g}_{abc}
\phi_a \phi_b \phi_c\,\,,
$$
for even parity couplings, and
$$
{\cal L}_{\rm int} = \sum_{abc} {1\over\sqrt{N}}\, {\hat g}_{abc}
\epsilon^{\mu\nu} \partial_{\mu}\phi_a \partial_{\nu}\phi_b \phi_c\,\,,
$$
for the odd parity couplings. Here $\phi_a$ represent meson
interpolating fields
and ${\hat g}_{nB\pi} = g_n (q^2 = \mu_n^2) / \mu_n^2 $. According
to the analysis in ref.~\refs\gmm, all the on-shell effective couplings
vanish in the chiral limit except the first one, $n=0$, which
approaches ${\hat g}_{BB\pi} \simeq 1$. It is interesting to note
though, that for non-zero pion mass the even and odd effective
couplings have quite different sizes. Because no extra
kinematical factor appears in the definition \gene\ for even parity
couplings, ${\hat g}_n$ for odd $n$ is typically ten times bigger
than ${\hat g}_n$ for even $n$.

The mechanism for the chiral supression is very interesting in
itself. For resonances above threshold, $\mu_n^2 > (\mu_B +
\mu_{\pi})^2 $, the on-shell coupling can be directly obtained from
the value of the vertex function, $\cal T$ at the point $\omega =
\omega (q^2 = \mu_n^2)$, as solved from  eq.~\onshell. In fig.~\rolling\
we localize this point on the $\cal T$ graph for two different light
quark masses. It is amusing to see that the source for the supression
is not the shrinking of the $\cal T$ function itself, but the
purely kinematical displacement of on-shell value of
$\omega$  so that the value of the on-shell coupling ``rolls down" to
the left of the vertex peak as we approach the chiral limit.
The vertex function itself actually grows in absolute value
as we set $m_q \rightarrow 0$. This means that non-local effects,
missed by the local expansion of an effective lagrangian, may become
very important for certain values of the parameters.
When computing  $1/N$
corrections in the full theory, it is the whole $\cal T$ function what
enters as integrand in loop diagrams, and we see that an estimate of the
corrections based on the size of the effective low-energy coupling
might not be correct, as compared to the complete non-local computation.

\newsec{A comment on analyticity}

When using the formulas of the preceeding section for $f_{\pm}(q^2)$
inside dispersion relations one should shift to the pole: $q^2
\rightarrow \mu_{n}^2$. Actually, the question of analyticity is a
subtle one in this formalism. Because we work in a
non-covariant gauge, in a singular frame (the light-cone quantization
surface), and with non-local effective interactions (the Coulomb
force), the analyticity of the off-shell Green's functions is not manifest.
In fact, as shown in refs.~\refs{\gross,\einh}, analyticity can be
directly verified in particular examples, although it follows from
rather non-trivial identities satisfied by the bound-state wave
functions.

{}From the point of view of our effective hamiltonian \ehff\
the problem is very clear.
Since we have vertices for $W$ emission by mesons,
but no possible $W$ decay  into two mesons, we see that
crossing symmetry is not manifest at tree level. To be more explicit,
consider, for example, the space-like form factor $F^{+} (q^2)$, defined
as the 1PI amplitude for $B$ decay into $W^+$ and $\pi$:
\eqn\ffs
{F^{+}(q^2)_{s-l} = C\Bigl\langle W^+ (q^+);\pi(P^+ -q^+)\Big\vert(q^- -
P_{0}^- ) {1\over q^- - P^{-}_{\rm eff} + i0}(q^- - P_{0}^- )\Big\vert
B(P^+)\Bigr\rangle_{1{\rm PI}}\,\,.  }
Now we have an extra ${\rO}(1)$ contribution to \ffs\ coming from
the direct vertex $B \rightarrow W^+ +\pi$ with the final result:
\eqn\qmodel
{\eqalign{F^+ (q^2)_{s-l}&= 2 q^+ \,{g^2 N\over\pi} \sum_n {f_n
\CT(B|B_n \, \pi;z) \over q^2 - \mu_{n}^2}  \cr
&+ 2P^+ \,(1-z)\int_{0}^{1} \d x\,\,\phi_B (z+(1-z)x)\,\phi_{\pi}(x)\,\,, }}
where $z\equiv q^+ /P^+$. So, in addition to the
superposition of resonances of the same form as in eq.~\dosf,
we get a quark
model-type contribution due to the bare emission of $W^+$ by the heavy
quark. (Formula \qmodel\ agrees with the Bethe-Salpeter analysis of
ref.~\refs\einh.)

As a consequence, eq.~\qmodel\ seems to favor a two-component
model for form factors (with bare, quark-model terms). In fact, there
is no contradiction between \qmodel\ and \dosf, due to the subtle
analytic properties of the three-vertex functions. We see that the
time-like kinematics,
$$
q^2 = {\mu_B^2 \over \omega} + {\mu_{\pi}^2 \over 1-\omega}\,\,, $$
and the space-like kinematics,
$$
\mu_B^2 = {q^2\over z} + {\mu_{\pi}^2 \over 1-z}\,\,, $$
are related by crossing: $B_n\leftrightarrow B$; $q^2
\leftrightarrow \mu_B^2$; $z\leftrightarrow \omega$. If we apply the
crossing transformation to the three-vertex
$\CT(B_n|B\,\pi;\omega)$ appearing in \dosf,
and analytically continue wave functions
through 't Hooft equation for $x\notin [0,1]$:
$$
\phi_n(x)=\left( {\mu_n^2\over g^2 N/\pi}
-{x_Q-1\over x}-{x_q -1\over1-x}\right)^{-1}\,\,
\pvi_{0}^{1} \d y\, {\phi_n (y) \over (y-x)^2}\,\,, $$
and further use the identity \ident\ we easily recover result \qmodel.

Thus, in this formalism, the two-component model picture and the pure
pole-dominance picture are complementary.
Direct computation of the decay in the real kinematical region $q^2
<(\mu_B - \mu_\pi)^2$ leads to eq.~\qmodel\ and it looks like a
two-component
model. However, one cannot relate directly the three-meson coupling
function $\CT(B|B_n \pi;z)$ appearing in \qmodel\ to the three-meson
coupling in a low-energy lagrangian because  $q^2 =\mu_n^2$
cannot be reached
with $z\in [0,1]$. When the appropriate analytic extension
to $q^2 > (\mu_B + \mu_\pi )^2$ is considered, the quark model-type
term is already included in the definiton of $\CT(B_n|B\pi;\omega)$,
which is directly related to the low-energy coupling when put on-shell,
$q^2 = \mu_n^2$. As it was pointed out by Jaffe and Mende~\refs\jaffe,
inside dispersion relations one must use a sum rule to compute the
residue at the poles outside the time-like kinematical region,
$\mu_n^2 < (\mu_B + \mu_\pi)^2$, because the interval
$(\mu_B - \mu_\pi )^2 <q^2 < (\mu_B +\mu_\pi)^2$ is not covered by
either of the definitions \ff\ or \ffs.

\newsec{Conclusions and outlook}

  In this paper we have discussed some aspects of the interplay
between the large-$N$ expansion and light-cone techniques in QCD.
Because of the special properties of light-cone Fock spaces
the large-$N$ bound-state problem reduces to
linear equations for the asymptotic bound states, without any need for
complicated combinatorial arguments involving Feynman diagrams.
Once the large-$N$ solution is known, the definition of effective many-body
hamiltonians renders the $1/N$ expansion
in terms of the asymptotic degrees of freedom completely systematic,
and provides a diagrammatic technique for the meson-glueball theory.

Although the general features of the method are largely independent
of the space-time dimension, quantitative results were reported only
for the case of two-dimensional QCD with fermions in the
fundamental representation. This is because the  effective
hamiltonians must be computed in terms of the large-$N$ solution, and
the 't Hooft model is the only QCD-like theory for which a complete
and simple parametrization of the large-$N$ physics is known.

 The main result of our analysis of $1/N$ corrections in QCD$_2$ is
that $1/N = 1/3$ is effectively a small parameter, at least for
moderately heavy quarks. The effective meson hamiltonian is weakly
coupled with $1/N$ corrections to spectrum and form factors of order a
few percent at most. However, the chiral limit looks rather singular
as far as $1/N$ corrections are concerned. In particular, it seems unlikely
that the chiral supression of higher resonances in $B$-meson form
factors, as discussed in ref.~\refs\gmm, could survive $1/N$
corrections in the strict chiral limit. This is in fact very natural
because Coleman's theorem forbids spontaneous symmetry breaking in the
interacting model. It is well known that the $N=\infty$ theory of
mesons is non-interacting, and it simulates a scenario of chiral
symmetry breaking, but we do not expect the two limits to commute.
Indeed, from our analysis of spectrum corrections in sect.~6 we find
that $1/N$ corrections renormalize the light quark masses in such a
way that the chiral symmetry point becomes unstable. As a result,
QCD$_2$ does not seem to be a good model for chiral dynamics beyond the
planar limit.

A general feature  of the light-cone effective hamiltonians
discussed here is that no low-energy expansion is required in
principle, not even in practice for the two-dimensional case.
In particular, we have pointed out some non-local effects for light
quark masses that could be underestimated by local effective
lagrangians. We also explained how some standard analyticity
properties are obtained in a very indirect way in this formalism, in
terms of non-trivial identities satisfied by the bound-state
wave functions.

An interesting point not developed in this paper is the formal
analogy between our effective hamiltonians and those of light-cone
string field theory. This analogy is based on both the structure of
the states in the Fock space representation, and also the perturbation
technique.  In ref.~\refs\bars\ it was shown that a Nambu-Goto
open string with massive ends, quantized in the light-cone gauge,
leads to 't Hooft bound-state equation as mass-shell condition. Thus,
we can conjecture that our effective meson hamiltonian provides the string
field theory construction for this string in the light-cone gauge.
However, 't Hooft wave functions enter our construction only as an
input. As a consequence, no further insights are provided towards a
string interpretation of QCD$_2$ in terms of the Gross-Taylor mapping
rules \refs\grosstay , which can be considered as the {\it ab initio}
definition of the pure Yang-Mills string.

Finally, regarding the four-dimensional generalization of these
techniques, we may say that any practical application of effective
light-cone hamiltonians    to the computation of hadron interactions
requires a good quantitave understanding of the large-$N$ bound-state
problem. Exact bound-state integral equations can be derived within
the light-cone method, and  the main simplification attained by the
large-$N$ limit is not a smaller hamiltonian, but a much simpler Fock space
structure. This fact, combined with the structure of the corrections
outlined in this paper, makes clear that the $1/N$ expansion
is a very interesting tool in the light-cone approach to QCD.

\bigskip\noindent
{\bf Acknowledgments}

We wish to thank D. Gross, I. Klebanov, P. Mende and H. Verlinde
for useful discussions. We are especially greatful to P. Mende
for his advice and help with numerical calculations.
The work of J.L.F.B. was supported by NSF Grant PHY90-21984.
The work of K.D. was supported in part by NSF Presidential Young
Investigator Award PHY-9157482 and James S. Mc.Donnell Foundation
Grant No. 91-48.

\appendix{A}{Vertex functions}

We collect here the explicit expressions for the vertex functions as
convolutions of wave functions. It is appropriate to define some
related useful quantities like the meson decay constant,
\eqn\fpi {f_n \equiv \int_{0}^{1} \d x\,\, \phi_n (x)\,\,,  }
the quark-antiquark-meson vertex,
\eqn\qqmv
{\Phi_n (x) \equiv \pvi_{0}^{1} \d y\,\, {\phi_n(y)\over (x-y)^2}\,\,, }
and the quark-antiquark-meson-meson vertex,
\eqn\qqmm
{\Psi_{n,m} (x;z,\omega) \equiv \pvi_{0}^{1} \d y \,\,{\phi_n (y) \phi_m
({z\over \omega} y) \over (x-y)^2}\,\,.   }

We also recall the action of the symmetry operators: the charge
conjugation operator,
$$\CC : \phi (x) \rightarrow \phi (1-x)\,\,,$$
the permutation operator on the right,
$$ \CP_R \equiv  (R_1 \leftrightarrow R_2 ; \omega \leftrightarrow
1-\omega )\,\,, $$
and similarly for $\CP_L$ acting on $L$-labels and $z$.

We can now write the triple-meson vertex functions:
\eqna\tvm
$$\eqalignno{
\CT(L|R_1 R_2 ; \omega )&= (1- \CC \CP_R )\,
{1\over 1-\omega} \int_{0}^{\omega} \d x \,
\phi_L (x) \phi_{R_1}\left( {x\over \omega}\right)\Phi_{R_2} \left({x-\omega
\over 1-\omega}\right)  \cr
&={1\over 1-\omega} \int_{0}^{\omega} \d x\, \phi_L (x) \,
\phi_{R_1}\left( {x\over\omega}\right)\, \Phi_{R_2}\left({x-\omega
\over 1-\omega}\right) &\tvm {\rm a} \cr
&- {1\over \omega}\int_{\omega}^{1} \d x\,\phi_L (x)
\,\phi_{R_2}\left({x-\omega \over
1-\omega}\right)\,\Phi_{R_1}\left({x\over\omega}\right)\,\,, }$$
and
$$\eqalignno{
\CT^{'} (L|R_1 R_2; \omega) &= (1-\CC)\, {f_{R_2}\over 1-\omega}
\int_{0}^{\omega} \d x \,\phi_L (x) \phi_{R_1} \left({x\over \omega}
\right)\,\,. &\tvm {\rm b} } $$
The four-meson vertices coming from the $BB+DD$ terms read:
\eqna\fmvf
$$\eqalignno {
\CF_{{\rm B}^2 +{\rm D}^2} &(L_1 L_2; z|R_1 R_2;\omega)=(1+ \CC\CP_R)
\Biggl[- \theta (\omega - z) {1-\omega \over z}
\int_{0}^{1} \d x\,
\phi_{L_2} \left({1-\omega \over 1-z}\,x \right)
\phi_{R_2} (x) \cr
&\times\Psi_{L_1 R_1} \left(1-(1-x)\,{1-\omega\over z};z,\omega\right)
+ (L\leftrightarrow R, \omega \leftrightarrow z) \Biggr]\,\,,
& \fmvf {\rm a} } $$
and
$$\eqalignno {
\CF^{'}_{{\rm B}^2 +{\rm D}^2}&(L_1 L_2 ; z|R_1 R_2;\omega)=(1+\CC )\,
\Biggl[-\theta (\omega - z)\, {z(1-\omega)\over (z-\omega)^2}
\int_{0}^{1} \d x\,
\phi_{L_1} (x)
\phi_{R_1} \left({z\over\omega}\,x\right) \cr
&\times  \int_{0}^{1}  \d y\,
\phi_{L_2}\left({1- \omega\over 1-z}\, y\right)
\phi_{R_2} (y)
+(L\leftrightarrow R, \omega\leftrightarrow z) \Biggr]\,\,.
&\fmvf {\rm b}  }$$
Finally, the four-point vertices from the $BD$ terms are given by:
\eqna\fmvff
$$\eqalignno {
\CF_{\rm BD} &(L_1 L_2 ; z|R_1 R_2 ; \omega) =
\Biggl[2 \theta (\omega - z)\, {1-\omega \over z}
\int_{0}^{1} \d x \,
\phi_{L_2}\left(1-{1-\omega\over 1-z}\,x\right)
\phi_{R_2}(1-x) \cr
&\times\Psi_{L_1 R_1} \left({\omega\over z}+(1-x)\,
{1-\omega\over 1-z}; z, \omega \right)
+(L\leftrightarrow R, \omega\leftrightarrow z) \Biggr]\,\,,
&\fmvff {\rm a} } $$
and
$$\eqalignno {
\CF^{'}_{\rm BD} &(L_1 L_2 ; z|R_1 R_2 ;\omega) = (1+ \CC)
\Biggl[ \theta (\omega -z)\,{z(1-\omega)\over (z-\omega)^2}
\int_{0}^{1} \d x\,
\phi_{L_1}(x)
\phi_{R_1}\left({z\over\omega}\,x\right) \cr
&\times\int_{0}^{1} \d y\,
\phi_{L_2}\left(1-{1-\omega\over 1-z}\, y\right)
\phi_{R_2} (1-y)
+(L\leftrightarrow R, \omega\leftrightarrow z) \Biggr]\,\,.
&\fmvff {\rm b} }$$

\listrefs
\vfil\eject
\bye